\documentclass[aps,prc,twocolumn,showpacs,preprintnumbers,
               nofootinbib,float,superscriptaddress]{revtex4-1}
\usepackage{graphicx, fancybox}
\usepackage{amsmath,amssymb}
\usepackage[colorlinks=true, pdfstartview=FitV, linkcolor=red,
citecolor=blue, urlcolor=blue]{hyperref}
\usepackage{color}
\usepackage{soul}
\usepackage{url}

\usepackage[normalem]{ulem}

\newcommand{\bdm}{\begin{displaymath}}
\newcommand{\edm}{\end{displaymath}}
\newcommand{\nn}{\nonumber}

\newcommand{\comment}[1]{}

\newcommand{\T}{\mathrm T}

\newcommand{\be}{\begin{equation}}
\newcommand{\ee}{\end{equation}}
\newcommand{\ba}{\begin{eqnarray}}
\newcommand{\ea}{\end{eqnarray}}

\newcommand{\gsim}{\mathrel{\hbox{\rlap{\lower.55ex \hbox {$\sim$}}
                   \kern-.3em \raise.4ex \hbox{$>$}}}}
\newcommand{\lsim}{\mathrel{\hbox{\rlap{\lower.55ex \hbox {$\sim$}}
                   \kern-.3em \raise.4ex \hbox{$<$}}}}

\def\be{\begin{eqnarray}}
\def\ee{\end{eqnarray}}

\def\roughly#1{\mathrel{\raise.3ex\hbox{$#1$\kern-.75em%
\lower1ex\hbox{$\sim$}}}}
\def\lsim{\roughly<}
\def\gsim{\roughly>}

\def\({\left(}
\def\){\right)}
\def\[{\left[}
\def\]{\right]}

\def\lsim{\mathrel{\rlap{\lower3pt\hbox{\hskip1pt$\sim$}}
     \raise1pt\hbox{$<$}}} 
\def\gsim{\mathrel{\rlap{\lower3pt\hbox{\hskip1pt$\sim$}}
     \raise1pt\hbox{$>$}}} 

\begin{document}

\title{The production of photons in relativistic heavy-ion collisions}

\author{Jean-Fran\c{c}ois Paquet}
\affiliation{Department of Physics, McGill University, 3600 University
Street, Montreal,
QC, H3A 2T8, Canada}
\affiliation{Department of Physics \& Astronomy, Stony Brook University, Stony Brook, NY 11794, USA}
 \author{Chun Shen}
 \affiliation{Department of Physics, McGill University, 3600 University
 Street, Montreal,
 QC, H3A 2T8, Canada}
\author{Gabriel S. Denicol}
 \affiliation{Department of Physics, McGill University, 3600 University
 Street, Montreal,
 QC, H3A 2T8, Canada}
\affiliation{Physics Department, Brookhaven National Laboratory, Upton, NY 11973, USA}
 
 \author{Matthew Luzum}
\affiliation{Departamento de F\'isica de Part\'iculas and IGFAE,
Universidade de Santiago de Compostela, E-15706 Santiago de Compostela, Galicia-Spain}
\affiliation{
Instituto de F\'isica - Universidade de S\~ao Paulo, Rua do Mat\~ao Travessa R, no. 187, 05508-090, Cidade Universit\'aria, S\~ao Paulo, Brasil
}

 \author{Bj\"orn Schenke}
 \affiliation{Physics Department, Brookhaven National Laboratory, Upton, NY
 11973, USA}
 
 \author{Sangyong Jeon}
 \affiliation{Department of Physics, McGill University, 3600 University
 Street, Montreal,
 QC, H3A 2T8, Canada}
 \author{Charles Gale}
 \affiliation{Department of Physics, McGill University, 3600 University
 Street, Montreal,
 QC, H3A 2T8, Canada}

\begin{abstract}
In this work it is shown that the use of a hydrodynamical model of heavy ion collisions which incorporates recent developments, together with updated photon emission rates, greatly improves agreement with both ALICE and PHENIX measurements of direct photons, supporting the idea that thermal photons are the dominant source of direct photon momentum anisotropy. The event-by-event hydrodynamical model uses IP-Glasma initial states and includes, for the first time, both shear and bulk viscosities, along with second order couplings between the two viscosities. The effect of both shear and bulk viscosities on the photon rates is studied, and those transport coefficients are shown to have measurable consequences on the photon momentum anisotropy.
\end{abstract}

\maketitle
\date{\today }

\section{Introduction}
The collision of heavy ions is the only way to produce and study hot and dense strongly interacting matter in the laboratory. Therefore, relativistic nuclear collisions represent a unique opportunity to explore QCD (Quantum ChromoDynamics) -- the theory of the strong interaction -- in extreme conditions of temperature and density.  A vibrant experimental program is currently under way at RHIC (the Relativistic Heavy Ion Collider, at Brookhaven National Laboratory, NY) and at the LHC (the Large Hadron Collider, at CERN, in Geneva). The data accumulated at these facilities represent unequivocal evidence that a new state of matter has been created in the collision of large nuclei: the Quark-Gluon Plasma (QGP) \cite{[{See, for example,  }][{, and references therein.}]Braun-Munzinger:2014pya}. One of the most striking features of the QGP is that its time-evolution can be modelled with relativistic fluid dynamics. Early studies concentrated on ideal fluids \cite{Kolb:2003dz,*Huovinen:2003fa}, but the realization that hadronic data from relativistic heavy ion collisions could be used to extract the transport coefficients of QCD -- in particular the shear viscosity to entropy density ratio, $\eta/s$ -- opened new horizons. More specifically, the azimuthal anisotropy of the particle momentum distribution in a single event  can be characterized by $v_n$, the coefficients of Fourier expansion in azimuthal angle $\phi$ \cite{Luzum:2013yya}: 
\be
E \frac{d^3 N}{d^3 p} &=& \frac{1}{2 \pi} 
 \frac{d N}{p_T d p_T d y}\left[1 + 2 \sum_{n=1}^\infty v_n \cos n \left(\phi - \Psi_n\right)\right], \nn \\
\ee
where $p_T$ is the transverse momentum,  
the $\Psi_n (p_T, y)$ are orientation angles, and $y$ is the rapidity. 
Viscous hydrodynamics calculations have established a quantitative connection between the empirically-extracted $v_n$'s and the value of $\eta/s$, the shear viscosity to entropy density dimensionless ratio \cite{Romatschke:2007mq,Teaney:2003kp}. 
Studies of the different experimental anisotropic flow coefficients have concluded that the phenomenologically-extracted $\eta/s$ values were close to the conjectured lower bound of $\eta/s = 1/4 \pi$ \cite{Kovtun:2004de}. 
The ability for fluid-dynamical models to quantitatively reproduce the measured behavior of the flow anisotropy coefficients of hadrons has been one of the major highlights of the entire relativistic heavy-ion program \cite{Jacak:2012dx}. 

A precise determination of quantities such as $\eta/s$ is made difficult by significant uncertainties in the description of the early time dynamics and the effect of additional sources of dissipation~\cite{Ryu:2015vwa}, among others. Electromagnetic observables produced in ultrarelativistic heavy ion collisions can be used as additional probes of the properties of the QGP, and can help constrain the transport coefficients of QCD.
The task of measuring photons and subtracting the large background of hadronic decay photons has been undertaken at RHIC~\cite{Adare:2008ab,Adare:2011zr,Adare:2014fwh,Adare:2015lcd} and the LHC~\cite{Wilde:2012wc,Lohner:2012ct,Adam:2015lda}, and the direct photon transverse momentum spectra and azimuthal anisotropy are available at both colliders. The observation that the magnitude of the direct photon $v_2$ was similar in size to that of hadrons, along with the exponential behaviour of the measured direct photon spectra at low transverse momentum, suggested that both photons and hadrons were produced by a similar mechanism: (quasi-)thermal production. 

While event-by-event hydrodynamical models of heavy ion collisions were shown repeatedly to provide a good description of hadronic observables~\cite{Gale:2013da, Heinz:2013th}, similar attempts at describing direct photon measurements did not meet with the same success \cite{Dion:2011pp,*Chatterjee:2013naa,*Shen:2013cca}. A simultaneous description of the direct photon spectra and momentum anisotropy proved to be a particular challenge.
In response to this apparent tension with measurements, investigations into additional photon production mechanisms multiplied~(e.g. \cite{McLerran:2015mda,Linnyk:2015tha,Tuchin:2014pka,Basar:2014swa,McLerran:2014hza,Gale:2014dfa,Monnai:2014kqa}). 

In this paper, a  hydrodynamical calculation of direct photon production is presented. It uses an up-to-date hydrodynamical model of heavy ion collisions~\cite{Ryu:2015vwa} along with the latest photon emission rates~\cite{Turbide:2003si,Heffernan:2014mla,Holt:2015cda}. Emphasis is put on photon emission from the expanding QCD medium, referred to as ``thermal photons''. 

The aim of this work is to present an up-to-date calculation of thermal photons using the latest developments in hydrodynamical simulation of heavy ion collisions. Understanding the current status of thermal photon production in heavy ion collisions will help guide future effort at identifying and constraining alternative photon production mechanisms.

\section{Hydrodynamical model}
\label{sec:hydro}

The relativistic fluid dynamics background that provides the time-dependent environment in which the photon-generation mechanisms evolve, is the same here as that used for hadrons in Ref. \cite{Ryu:2015vwa}. We summarize its main features again here, for convenience. The initial state of the nuclear collision are modelled using the IP-Glasma approach \cite{Schenke:2012wb}, which builds on the  ``impact parameter dependent saturation model''  (IP-Sat) \cite{Bartels:2002cj,Kowalski:2003hm} that constrains the distribution of initial colour sources drawing from electron-proton and electron-nucleus collision data. The gluon fields are then evolved in space and time using classical Yang-Mills equations: $\left[D_\mu,F^{\mu \nu}\right] = 0$, up to a proper time $\tau_0$ of order of the inverse of the saturation scale. 
The energy density, $\epsilon$, and the flow velocities, $u^\mu$, from the Yang-Mills evolution are then used to initialize the hydrodynamical evolution. This is achieved by solving $u_\mu(\tau_0)T_{\textrm{CYM}}^{\mu \nu} (\tau_0) = \epsilon (\tau_0) u^\nu (\tau_0)$ where $T_{\textrm{CYM}}^{\mu \nu}$ is the classical Yang-Mills energy-momentum tensor. As in Ref.~\cite{Ryu:2015vwa}, $\tau_0 = 0.4$~fm is used in this work. The IP-Glasma initial conditions are boost-invariant, and the subsequent hydrodynamical evolution is $2+1$D as well.

\begin{figure}[tb]
 \begin{center}
                \includegraphics[width=0.46\textwidth]{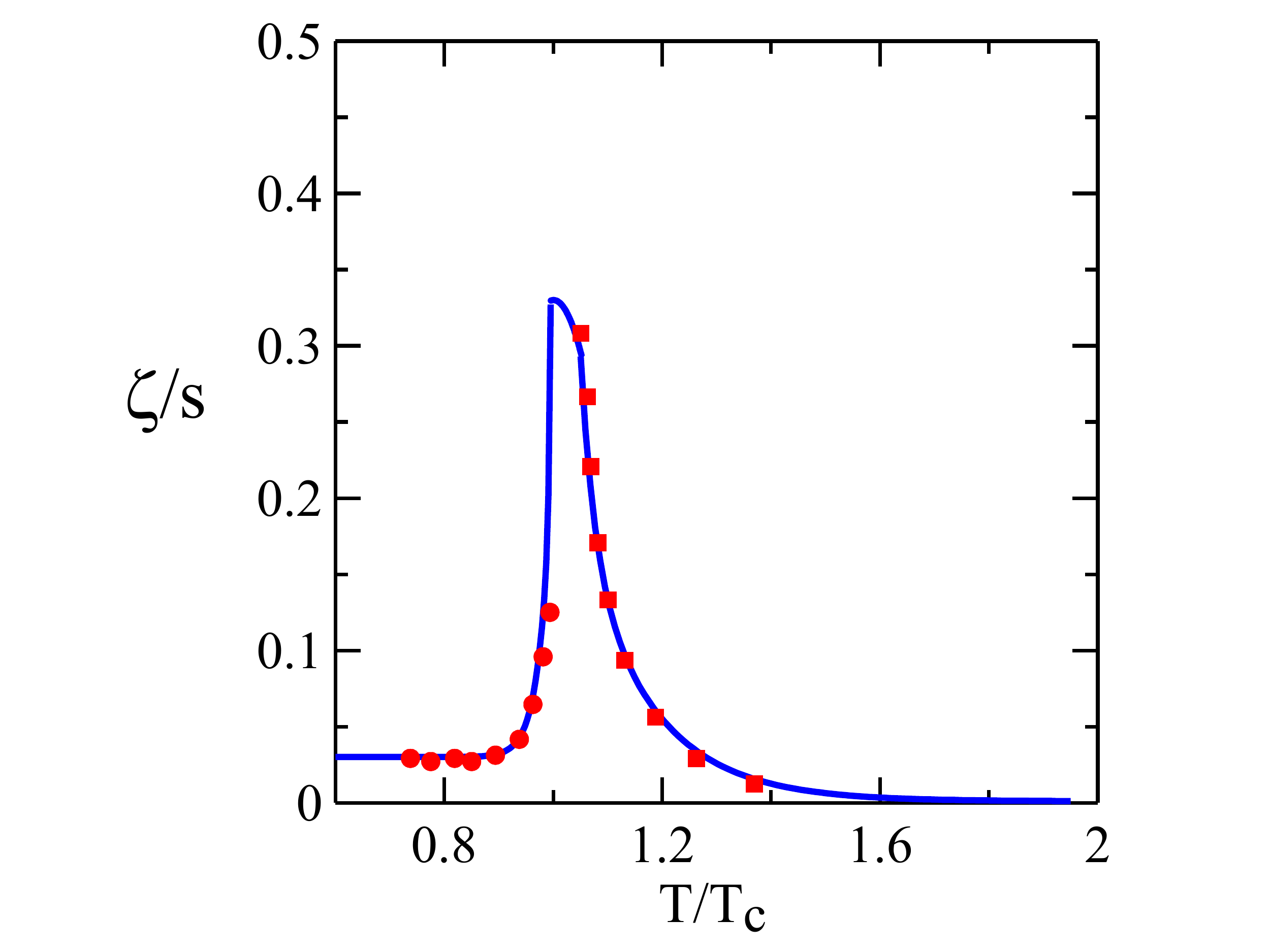}
\end{center}
\vspace*{-0.5cm}
\caption{The temperature dependence of the bulk viscosity to entropy density ratio, as used in this work. The points on the low temperature side are the results of a calculation from Ref.~\cite{NoronhaHostler:2008ju}, while the high-temperature results are from Ref.~\cite{Karsch:2007jc}. }
\label{fig:bulk}
\end{figure}

The hydrodynamic evolution involves a dissipative term in the stress-energy tensor:
\begin{eqnarray}
\T^{\mu \nu}_{\rm diss} &=& \pi^{\mu \nu} - \Delta^{\mu \nu} \Pi \;,
\end{eqnarray}
where $\Delta^{\mu \nu} = g^{\mu \nu} - u^\mu u^\nu$.
In the above, $g^{\mu \nu}=diag(1,-1,-1,-1)$ is the Minkowski tensor, $\pi^{\mu \nu}$ is the shear-stress tensor, and $\Pi$ is the bulk pressure term. The time-evolution of these last two quantities is obtained by solving relaxation-type equations \cite{Denicol:2012cn,Denicol:2014vaa}:
\begin{align}
\tau_{\Pi }\dot{\Pi}+\Pi = -\zeta \theta -\delta _{\Pi \Pi }\Pi \theta
+\lambda _{\Pi \pi }\pi ^{\mu \nu }\sigma _{\mu \nu }\;,  \label{intro_1}
\\
\tau_{\pi }\dot{\pi}^{\left\langle \mu \nu \right\rangle }+\pi ^{\mu \nu }
= 2\eta \sigma ^{\mu \nu }-\delta _{\pi \pi }\pi ^{\mu \nu }\theta
+\varphi
_{7}\pi _{\alpha }^{\left\langle \mu \right. }\pi ^{\left. \nu
\right\rangle
\alpha } \notag \\
-\tau _{\pi \pi }\pi _{\alpha }^{\left\langle \mu \right. }\sigma
^{\left. \nu \right\rangle \alpha }+\lambda _{\pi \Pi }\Pi \sigma ^{\mu
\nu
},
\label{eq:relax}
\end{align}%
with the definition
\begin{equation*}
A^{\langle \mu \nu \rangle }=\Delta _{\alpha \beta}^{\mu \nu }A^{\alpha \beta },
\end{equation*}
where 
\begin{equation*}
\Delta _{\mu \nu }^{\alpha \beta }=\frac{1}{2}\left[
\Delta _{\alpha }^{\mu }\Delta _{\beta }^{\nu }+\Delta _{\alpha }^{\nu
}\Delta _{\beta }^{\mu }-\frac{2}{3}\Delta ^{\mu \nu }\Delta _{\alpha \beta }%
\right]
\end{equation*}
 is the double, symmetric, and traceless projection operator. The expansion rate of the fluid is $\theta =\partial _{\mu }u^{\mu }$ and the shear tensor $\sigma ^{\mu \nu }=\partial ^{\langle \mu}u^{\nu \rangle }$. In the present work, the shear-stress tensor and the bulk pressure are initialized to zero at time $\tau_0$.

The second-order transport coefficients $\tau _{\Pi }$, $\delta _{\Pi \Pi }$, $\lambda _{\Pi \pi }$, $\tau _{\pi }$, $\eta $, $%
\delta _{\pi \pi }$, $\varphi _{7}$, $\tau _{\pi \pi }$, and $\lambda_{\pi\Pi }$ are related to the shear viscosity $\eta$ and bulk viscosity $\zeta$ using formulae derived from the Boltzmann equation near the conformal limit~\cite{Denicol:2014vaa}.

Importantly, the hydrodynamic evolution stage is followed by a phase where discrete particles are produced through the Cooper-Frye procedure \cite{Cooper:1974mv}. Late stage hadrons further interact and freeze-out dynamically through the UrQMD approach and algorithms \cite{Bass:1998ca}. 

Since viscous hydrodynamics is used, the medium is not exactly in thermal equilibrium. Consequently, whenever specific particle distributions are invoked -- in the Cooper-Frye scheme or for thermal photon production -- these will receive viscous corrections.
In the present work, the momentum distribution $f_{B/F}(P,X)$ is derived in Appendices (\ref{appendixA}) and (\ref{appendixB}), and is given by
\be
f_{B/F}(P,X)=f_{B/F}^{(0)}(P)+\delta f_{B/F}^{\rm shear}(P,X)+\delta f_{B/F}^{\rm bulk}(P,X)\;,\nn\\
\label{eq:fHadrons}
\ee
where
\be
\delta f_{B/F}^{\rm shear}(P,X)=f_{B/F}^{(0)}(P) (1 + \sigma_{B/F} f^{(0)}(P)) \frac{\pi^{\mu\nu} P^\mu P^\nu}{2 T^2 (\epsilon+\mathcal{P})}\nn\\
\label{eq:fShearHadrons}
\ee
and
\ba
\delta f_{B/F}^{\rm bulk}(P,X)&&=- f_{B/F}^{(0)}(P) (1 + \sigma_{B/F} f^{(0)}(P) ) \nn\\
&& \times \left[ \frac{1}{3} \frac{m^2}{T^2} \frac{1}{P\cdot u/T}-\frac{P\cdot u}{T} \left( \frac{1}{3}-c_s^2 \right) \right] \Pi \frac{\tau_\Pi}{\zeta} \nn \\
\label{eq:fBulkHadrons}
\ea
with $\sigma_{B}=1$ for bosons and $\sigma_{F}=-1$ for fermions, with $f_{B/F}^{(0)}(P)$ being correspondingly either the Fermi-Dirac or Bose-Einstein distribution. The pressure $\mathcal{P}$, energy density $\epsilon$, flow velocity $u$, speed-of-sound $c_s$ and temperature $T$ entering into the distribution functions are evaluated at spacetime point $X$. The momentum of the quasi-particle of mass $m$ is denoted $P$. The bulk relaxation time is $\tau_\Pi$.

Within the hybrid approach used here (IP-Glasma -- dissipative hydrodynamics -- UrQMD) a recent analysis of ALICE and CMS measurements indicates that identified particle spectra, multiplicity distributions, and multiple flow coefficients ($v_n \{2\}$, $n=2, 3, 4$) can be globally reproduced, for pions, kaons, and protons \cite{Ryu:2015vwa}. The calculation and data analyses were done for 0 - 5\% through 30 - 40\% centrality classes. These  LHC-energy analyses, which  include both bulk and shear viscosity, lead to $\eta/s=$ 0.095. A relatively narrow $\zeta/s$ temperature-profile, illustrated in Fig. \ref{fig:bulk}, was used. This profile peaks such that $\zeta/s (T_{\rm peak}) \sim 0.3$, with $T_{\rm peak} = 180$ MeV. As a reminder, this is a somewhat novel feature that most fluid-dynamical approaches to the modelling of relativistic heavy-ion collisions do not yet include. The approach used here also notably features non-linear terms that couple the shear and bulk sectors of the viscous hydrodynamics~\cite{Denicol:2014vaa}. 

Describing the late stage dynamic of the medium with an afterburner leads to significant complications in the evaluation of photon emission. As a consequence, late photon emission is not evaluated with the afterburner, but rather with hydrodynamics. The exact approach used is explained later in this work.

\section{Photon sources}
\label{sec:photonSources}

The photons measured in relativistic nucleus-nucleus collisions come from a variety of different sources, and these will be discussed in turn in this section. They fall in two broad categories: those with a thermal origin and those coming from ``cold'' processes. In  searches for signals from the quark-gluon plasma, the latter are usually thought of as a background. Since this background is irreducible in the experimental measurements, they nevertheless deserve our full attention. We start by describing how prompt photon production is evaluated in this work.

\subsection{Prompt photons}

In the very first instants of the nuclear collision, the interacting nucleons will produce photons through partonic Compton interactions and quark-antiquark annihilations. In addition, QCD jets will be generated and these jets will fragment into many final states, some of which will include photons. 

\begin{figure}[tb]
                \includegraphics[width=0.35\textwidth]{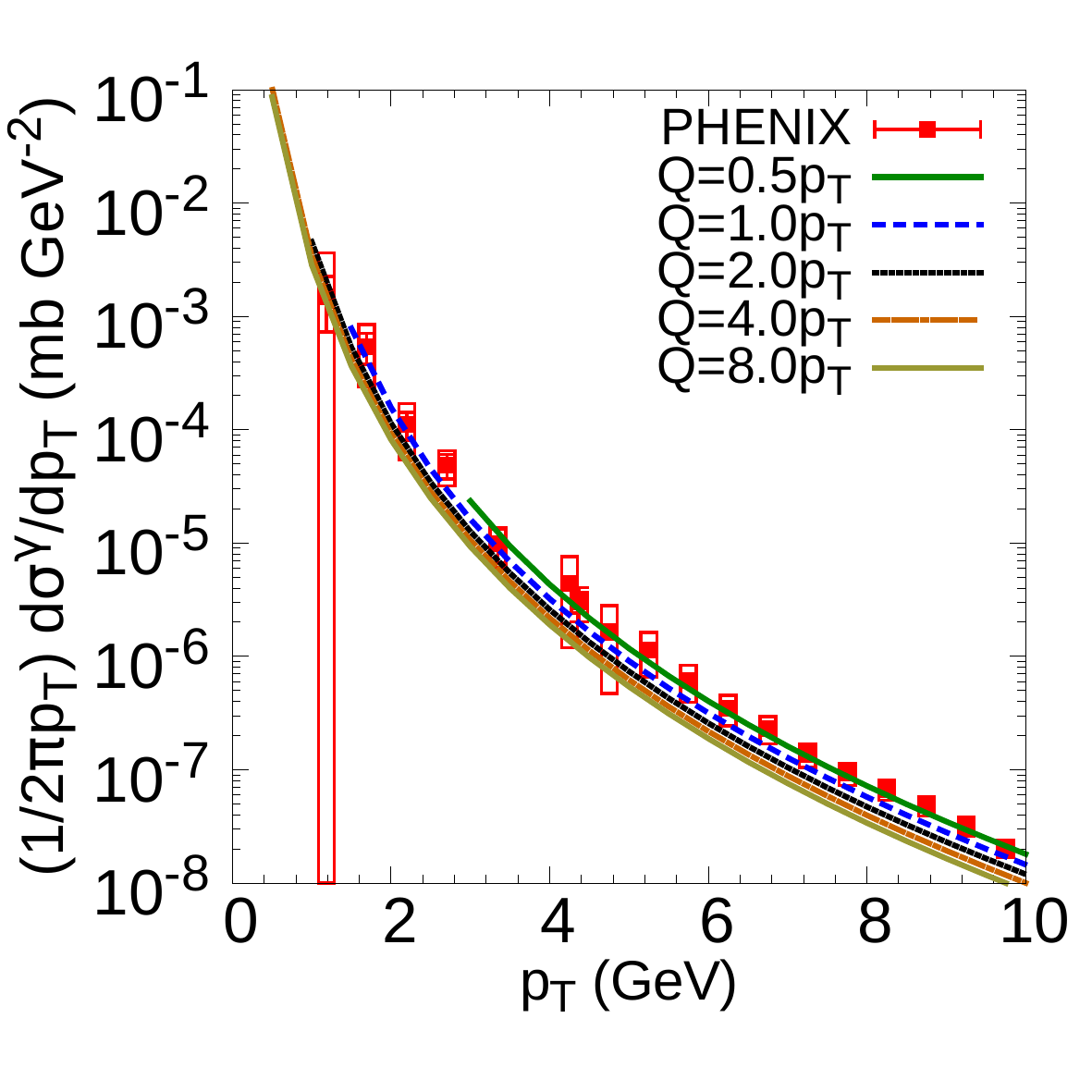}
                \includegraphics[width=0.35\textwidth]{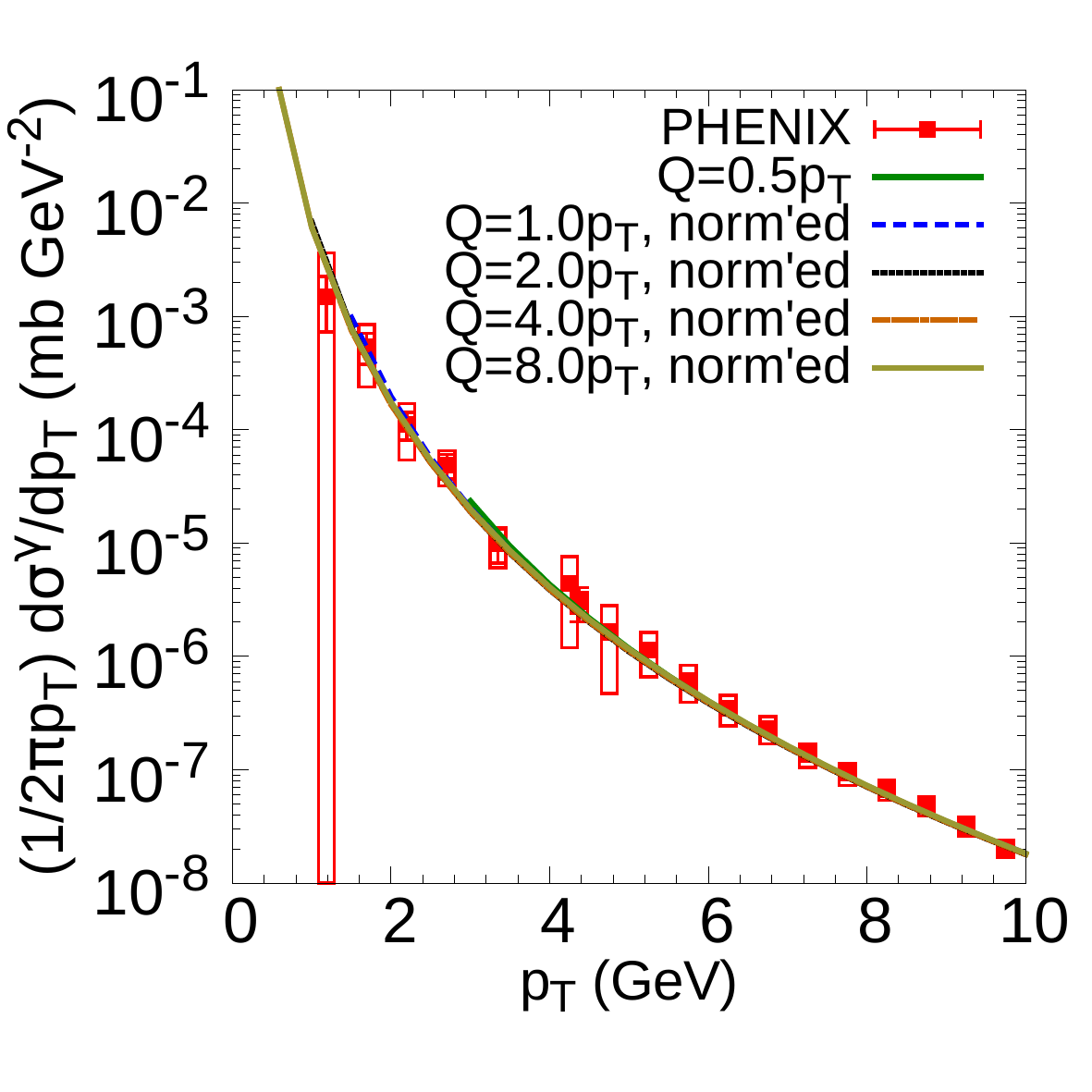}
\caption{Top panel: Direct photon spectrum measured in $\sqrt{s_{NN}}=200$~GeV proton-proton collisions at RHIC compared with  perturbative QCD calculations at different scales $Q$. Bottom panel: normalised perturbative QCD calculations; see details in the main text. }
\label{fig:ppRHICphotons}
\end{figure}

The calculation of photon production in hadronic interactions using the techniques of perturbative QCD has a long history \cite{[{See, for example, }][{, for an early review.}]Owens:1986mp} which has led to a fairly mature understanding of the subject. The photon production cross section in proton-proton collisions can be written concisely  as
\begin{eqnarray}
E \frac{d^3 \sigma_{\rm p p }}{d^3 p} = \sum_{a, b, c, d} f_{a/p} \left(x_a,Q_{\textrm{fact}}\right) 
\otimes  f_{b/p}\left(x_b,Q_{\textrm{fact}} \right) \nonumber \\ \otimes \, d\hat{\sigma}\left(Q_{\textrm{ren}}\right) \otimes D_{\gamma/c} \left(z_c, Q_{\textrm{frag}}\right),
\end{eqnarray}
where $Q_{\textrm{fact}}, Q_{\textrm{ren}}$, and $Q_{\textrm{frag}}$ are energy scales entering respectively into the parton distribution function $f_a$, the partonic cross-section $d \hat{\sigma}$, and the fragmentation function $D_{\gamma/c}$. The cross-section $d\hat{\sigma}(Q_{\textrm{ren}})$ is evaluated as a perturbative expansion in the strong coupling constant $\alpha_s(Q)$, and $Q_{\textrm{ren}}$ is the scale at which $\alpha_s(Q)$ is evaluated. This scale, along with the factorisation and fragmentation scales, should typically be of the order of the transverse momentum of final state partons. 

\begin{table*}[htb]
\begin{tabular}{|c|p{2cm}|p{2cm}|p{2cm}|p{2cm}|}
\hline
 & \multicolumn{2}{|c|}{RHIC Au-Au $\sqrt{s_{N N}}=200$~GeV} & \multicolumn{2}{|c|}{LHC Pb-Pb $\sqrt{s_{N N}}=2760$~GeV} \\
\hline
Centrality & 0-20\% & 20-40\% & 0-20\% & 20-40\% \\
\hline
$\langle N_{\textrm{coll}} \rangle$ & 793 & 323 & 1231 & 501 \\
\hline
\end{tabular}
\caption{Average number of binary collisions in different centrality classes, at RHIC and the LHC. Nucleon positions sampled from a Wood-Saxon distribution, which are input of the IP-Glasma model, are used to evaluate the number of binary collisions in each event with the MC-Glauber model. Centralities are defined using the gluon multiplicity of each event, as described in Ref~\cite{Gale:2012rq}.}
\label{table:Ncoll}
\end{table*}

Computing photon production in proton-proton collisions from the formalism described above requires a proton parton distribution function $f_{a/p} \left(x_a,Q_{\textrm{fact}}\right)$, a parton-to-photon fragmentation function $D_{\gamma/c}\left(z_c, Q_{\textrm{frag}}\right)$, and the partonic cross-section $d\hat{\sigma}\left(Q_{\textrm{ren}}\right)$. The latter is currently known at next-to-leading order in the strong coupling constant for both isolated photons \cite{Aurenche:1987fs} and fragmentation photons \cite{Aversa:1988vb}. Combined with next-to-leading order parton distribution and fragmentation functions, photon production using perturbative QCD has been shown  to agree very well with direct photon measurements in proton-proton collisions at RHIC, at the LHC,  and at previous colliders~\cite{Aurenche:2006vj}.

At high $p_T^\gamma$, prompt photons are by far the dominant source of direct photons. Those are calculated in heavy ion collisions by multiplying the  number of photons produced in proton-collisions by the number of binary collisions~\cite{Chatrchyan:2012vq,Afanasiev:2012dg}.
The scaling procedure may be applied either to a fit of direct photon measurements in proton-proton collisions, or to a perturbative QCD calculation of photon production. The latter is used in this work, 
in conjunction with nuclear parton distribution functions EPS09~\cite{Eskola:2009uj}, which take into account cold nuclear matter effects.
The study of fragmentation photon energy loss and jet-medium photon productions, two effects that are understood to modify low $p_T^\gamma$ prompt photon production in heavy ion collisions, is not undertaken here and will be the subject of a separate work.

Prompt photons are added to other sources of direct photons on an event-by-event basis. The perturbative QCD calculation is thus scaled by the number of binary collisions in each event individually. For reference, we quote in Table~\ref{table:Ncoll} the centrality-averaged number of binary collisions in each centrality studied in this work.

The pQCD framework used in this work is essentially the next-to-leading order calculation contained in the numerical code INCNLO \cite{incnlo}. The proton parton distribution function and photon fragmentation function used are respectively CTEQ61m \cite{Stump:2003yu} and BFG-2 \cite{Bourhis:1997yu}. The factorisation, renormalisation and fragmentation scales are all set equal to each other: $Q_{\textrm{fact}} = Q_{\textrm{ren}} = Q_{\textrm{frag}} = Q$. 
The transverse momentum of the produced photon is used to set the scale, with a normalization constant $Q = \lambda p^\gamma_T$. 
The effect of changing the proportionality constant between $Q$ and $p_T^\gamma$ is essentially a change in the normalization of the prompt photon spectra. This can be seen in Fig.~\ref{fig:ppRHICphotons}. The top panel shows the perturbative calculation of prompt photons in proton-proton collisions for different choices of scale $Q$, from $Q=p_T^\gamma/2$ to $Q=8 p_T^\gamma$. The lower panel shows the same calculations scaled by a constant so that they have the same normalization. It is clear from this last figure that the calculations overlap very well, showing that they have the same $p_T^\gamma$ dependence. It was verified that changing the scale $Q$ at LHC energies also has the same effect on the photon spectrum, i.e. it changes the normalisation but not the transverse momentum dependence of the calculation.

It is apparent that a small proportionality constant between $Q$ and $p_T^\gamma$, such as $Q=p_T^\gamma/2$, provides a better description of the available measurements at RHIC. On the other hand, it can be seen in Fig.~\ref{fig:ppRHICphotons} that calculations are limited to $p_T^\gamma>(1.5~$GeV$)/\lambda$, where $\lambda$ is the proportionality constant between $Q$ and $p_T^\gamma$. This limitation results from the presence of a scale $Q_0\sim 1.5$~GeV, which is typically taken as the limit of applicability of perturbative QCD. Parton distribution functions and fragmentation functions are usually limited to $Q>Q_0\sim 1.5$~GeV. Calculations made with INCNLO are also subject to this limit in $Q$. Although this might appear to limit the value of $p_T^\gamma$ at which prompt photons can be evaluated with perturbative QCD, the scaling behaviour observed on the bottom panel of Fig.~\ref{fig:ppRHICphotons} shows that it is not the case:  the effect of the scale $Q$ is simply a change in normalization, and prompt photons can be evaluated at low $p_T^\gamma$ by using e.g. $Q=4 p_T^\gamma$ and changing the normalization of the calculation to that of $Q=p_T^\gamma/2$. This is the procedure adopted in this work.

It remains that pQCD is based on the idea that a large momentum exchange occurs in a hadronic collisions, allowing for a part of the cross-section to be computed perturbatively. It is understood that perturbative QCD eventually breaks down at low transverse momentum, although the exact value of $p_T^\gamma$ at which this happens is not clear. As discussed above, the perturbative QCD calculation of prompt photons is in good agreement with the low $p_T^\gamma$ direct photon measurements in proton-proton collisions, although it slightly overestimates the very lowest point around $1$~GeV. It remains to be seen how much lower in transverse momentum the agreement with data persists.

While there are currently no low $p_T^\gamma$ photon data from proton-proton collisions at the LHC, the same extrapolation procedure was used to evaluate low $p_T$ $\pi^0$ production as an additional verification of the approach. Good agreement with measurements was again found down to $p_T\sim 1-2$~GeV~\cite{Paquet:2015Thesis}.

It is worth noting that, besides clarifying the domain of validity of perturbative QCD calculation of prompt photons, additional measurements will also help constrain uncertainties due to the photon fragmentation function, which are significant in the soft domains of perturbative QCD calculation of prompt photons \cite{Klasen:2013mga,*Klasen:2014xfa}. Without direct measurements, those uncertainties will likely persist.

\subsection{Thermal photons}
\label{sec:thermal}

The ``thermal photons'' are those photons resulting from the interaction of  thermalized medium constituents\footnote{The thermalization approximation will be relaxed later.}. The computation of photon production rates may be done using thermal field theory techniques, or using relativistic kinetic theory \cite{[{See, for example, }][{, and references therein.}]Kapusta:2006pm}. Both approaches have contributed to the  compendium of rates used in this work.

In the partonic sector, photon-production processes calculated at leading order in the strong coupling constant, $g_s$, have been available for almost 15 years \cite{Arnold:2001ms}. Those are used here\footnote{Some recent work has extended this seminal result by going up next-to-leading order \cite{Ghiglieri:2013gia}. For values of the strong coupling relevant to the phenomenology considered in the current work,  the net photon rate at NLO is a modest 20\% larger than that at LO.}. At high energies, the charged particle multiplicity is dominated by mesons. In the hadronic sector at temperatures comparable to, and lower than, the crossover temperature, photons originating from thermal reactions of mesonic origins were calculated in Ref. \cite{Turbide:2003si}. That same work also includes the photons obtained from taking the $\rho$-meson self-energy to zero invariant mass. This procedure accounts for the baryonic contributions, be it radiative decays or reactions of the type $\pi N \to \pi N \gamma$, and $N N \to N N \gamma$, where $N$ represents a nucleon. The net rate parametrized in Ref.~\cite{Turbide:2003si} also avoids  possible double-counting issues between mesonic and baryonic contributions. Finally, this work includes also  recent estimates of $\pi \pi$ bremsstrahlung contributions \cite{Heffernan:2014mla}, and of the reactions $\pi \rho \to \omega \gamma$, $\pi \omega \to \rho \gamma$, and $\pi \omega \to \rho \pi$ \cite{Holt:2015cda}, absent from Ref. \cite{Turbide:2003si}.
\begin{figure}[tb]
\centering
\includegraphics[width=0.7\linewidth]{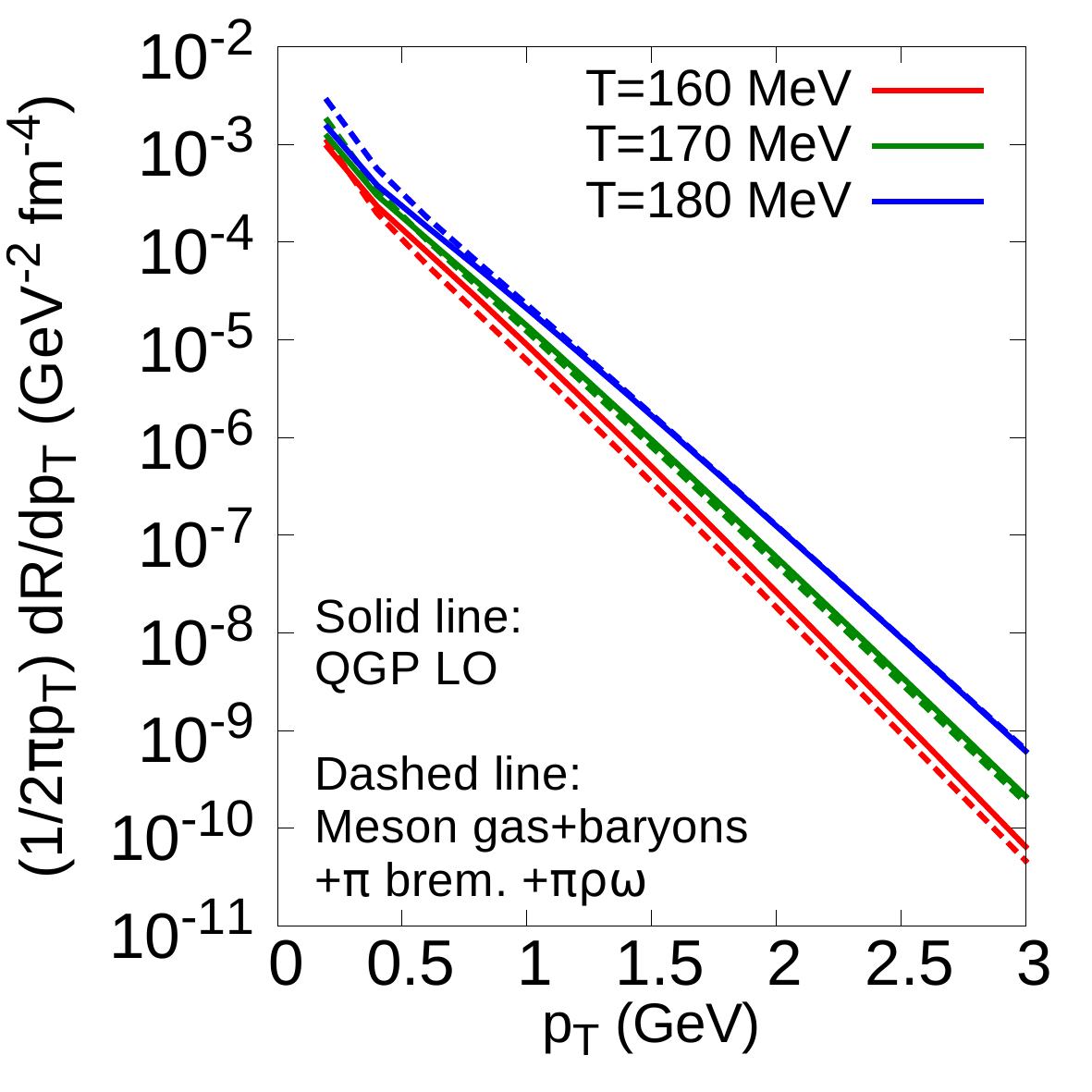}
\caption{Ideal QGP and hadronic photon rate near the cross-over region.}
\label{fig:rates}
\end{figure}
It is instructive to compare rates, prior to integrating them with a dynamical four-volume evolution. This is done in Fig.~\ref{fig:rates}. The figure shows the LO partonic rates of Ref. \cite{Arnold:2001ms} (solid lines) compared with the hadronic rates of Refs. \cite{Turbide:2003si,Heffernan:2014mla,Holt:2015cda} (dashed lines) for a range of temperatures in the cross-over region.

\subsection{Non-cocktail hadronic decay photons}
\label{sec:nonCocktail}

As the strongly-interacting fluid hadronizes, it transforms into hadrons which will interact. When those interactions cease, the momentum distributions are frozen and the particles free-stream out to the experimental detectors. The longer-lived hadrons will contribute significantly to the photon signal and therefore have to be included. Collectively, they are dubbed ``the cocktail'' and are (for ALICE) $\pi^0, \eta, \rho, \omega, \eta',\phi$; the relevant photon-producing decays are subtracted from the measured inclusive signal \cite{Lohner}, to expose a combination of  thermal photons and prompt photons. There are however other, shorter-lived, states which decay with a photonic component in the final states \cite{Agashe:2014kda}. This work includes all of the ones with a mass $M < 1.7$ GeV. The differential cross section of the decay photons can then be calculated, knowing the relevant branching ratio. After including all of these, together with the decays considered in Ref. \cite{Rapp:1999qu}, the most important channels were found to be $\Sigma \to \Lambda \gamma$, $f_1 (1285) \to \rho^0 \gamma$, and $K^*(982) \to K \gamma$. All contributions are however included, for completeness.

\section{Correcting the photon emission rates for viscosity}
\label{sec:visc_rates}

As mentioned earlier, it is an established fact that the bulk dynamics of strongly interacting matter is sensitive to the value of shear and bulk viscosities, two of the transport coefficients of QCD. Switching to a corpuscular description, and considering separately the reactions that, together, define the fluid enables a channel-by-channel viscous correction of the photon emission rates. The photon production rate, $R_\gamma$, admits a kinetic theory formulation. For $2 \to 2$ scattering $(1 + 2 \to 3 + \gamma)$ it is \cite{Kapusta:2006pm}
\ba
&&\omega \frac{d^3 R_{\gamma}}{d^3 k}=\frac{1}{2(2\pi)^3} \int \frac{d^3 p_1}{2 P^0_1 (2\pi)^3} \frac{d^3 p_2}{2 P^0_2 (2\pi)^3} \frac{d^3 p_3}{2 P^0_3 (2\pi)^3} \nn \\
&& \times  (2\pi)^4\delta^4(P_1+P_2-P_3-K) |\mathcal{M}|^2 
 f_{B/F}(P_1) f_{B/F}(P_2)\nn\\&&\times \left(1+\sigma_{B/F} f_{B/F}(P_3)\right),
\label{eq:photonProductionKin}
\ea
where $|\mathcal{M}|^2$ is the squared matrix element corresponding to the $2\to 2$ scattering, $f_{B/F}$ is the particle momentum distribution for bosons ($\sigma_{B}=1$) or fermions ($\sigma_{F}=-1$), and the photon four-momentum is $K = (\omega, \vec{k})$. The distribution function must then be modified in the presence of dissipative effects, for the kinetic formulation of $T^{\mu \nu}$ to match that with the explicit dissipative transport coefficients. This modification is written as $f_{B/F} = f_{B/F}^{(0)} + \delta f_{B/F}$. 

Linearizing in $\delta f_{B/F}$ yields
\ba
\omega \frac{d^3 R_{\gamma}}{d^3 k}\approx \omega \frac{d^3 R_{\gamma}^{(0)}}{d^3 k} + \omega \frac{d^3 R_{\gamma}}{d^3 k}^{\rm (visc)},
\ea
where
\ba
&&\omega \frac{d^3 R_{\gamma}}{d^3 k}^{\rm (visc)}  = \frac{1}{2(2\pi)^3} \int \frac{d^3 p_1}{2 P^0_1 (2\pi)^3} \frac{d^3 p_2}{2 P^0_2 (2\pi)^3} \frac{d^3 p_3}{2 P^0_3 (2\pi)^3}\nn\\
&& \times  (2\pi)^4\delta^4(P_1+P_2-P_3-K) |\mathcal{M}|^2 \nn\\
&& \times  \left[ \delta f_{B/F}(P_1) f^{(0)}_{B/F}(P_2) \left(1+\sigma_{B/F} \,f^{(0)}_{B/F}(P_3)\right) \right. \nn\\
& &  \qquad + f^{(0)}_{B/F}(P_1) \delta f_{B/F}(P_2) \left(1+\sigma_{B/F} \,f^{(0)}_{B/F}(P_3)\right) \nn \\
& & \left. \qquad  + f^{(0)}_{B/F}(P_1) f^{(0)}_{B/F}(P_2) \left(\sigma_{B/F} \,\delta f_{B/F}(P_3)\right) \right] .
\label{eq:photonProductionKinVisc}
\ea
Next, the corrections appropriate for shear and bulk viscosity are discussed.
\begingroup
\begin{table*}[ht]
\begin{tabular}{|c|c|c|c|}
\hline
Rate & Ideal & Shear correction & Bulk correction \\
\hline
QGP --- $2\to 2$ &  \cite{Arnold:2001ms} & Yes \cite{Shen:2014nfa} & Forward scattering approximation\\
\hline
QGP --- Bremsstrahlung &  \cite{Arnold:2001ms} & No & No \\
\hline
Hadronic --- Meson gas ($\pi$, $K$, $\rho$, $K^*$, $a_1$) &  \cite{Turbide:2003si} & Yes \cite{Dion:2011pp,Shen:2014thesis} & Yes [this work] \\
\hline
Hadronic --- $\rho$ spectral function (incl. baryons) &  \cite{Turbide:2003si,Heffernan:2014mla} & No & No \\
\hline
Hadronic --- $\pi+\pi$ bremsstrahlung &  \cite{Liu:2007zzw,Heffernan:2014mla} & No & No \\
\hline
Hadronic --- $\pi$-$\rho$-$\omega$ system &  \cite{Holt:2015cda} & No & No \\
\hline
\end{tabular}
\caption{A summary of the thermal photon rates sources, together with the current state of advancement of their viscous correction. The bulk corrections are original to this work. The ``forward scattering'' approximation refers to considering cases where, in $2 \to 2$ scattering, the exchanged momentum is soft (i.e.  $\sim gT$ ). In this case the amplitude will be dominated by forward scattering.}
\label{table:rates}
\end{table*}
\endgroup
While there is currently a considerable body of research devoted to the extraction of the shear viscosity from relativistic heavy-ion collisions \cite{[{See, for example, }][{, and references therein}]Gale:2013da}, studies of the bulk viscosity are rarer. To proceed further, assumptions need to be made about the form of $\delta f_{B/F}(P,X)$. The space-time coordinates of the emission site, $X$, are now explicit. The first-order correction $\delta f_{B/F}(P,X)$ is linear in the shear stress tensor~$\pi^{\mu\nu}$ and the bulk pressure~$\Pi$. In this case:
\be
\delta f_{B/F}(P,X)= \pi_{\mu\nu}(X) P^\mu P^\nu S(P,X) + \Pi(X) B(P,X), \nn\\
\ee
where two properties of $\pi_{\mu\nu}(X)$, $\pi_{\mu\nu}(X) g^{\mu\nu}=0$ and $\pi_{\mu\nu}(X) u^{\mu}=0$, were used to constrain the expansion of $\delta f_{B/F}(P,X)$ in $\pi_{\mu\nu}(X)$.

The functions $S(P,X)$ and $B(P,X)$ can depend on the spacetime position $X$ through e.g. the local value of the temperature $T(X)$, the energy density $\epsilon(X)$, the entropy density $s(X)$, etc. All these implicit functions of $X$ are thermodynamical quantities that are related through the equation of state of the medium. For practical reasons, it is better if corrections to photon emission do not have an explicit dependence on the equation of state. This can be achieved if the momentum dependence of $S(P,X)$ and $B(P,X)$ can be factorised from the rest such as:
\ba
\delta f_{B/F}(P,X)&=& \pi_{\mu\nu}(X) P^\mu P^\nu \sum_j S_X^{(j)}(X) S_M^{(j)}(P,T) \nn\\
&& +\, \Pi(X) \sum_j B_X^{(j)}(X) B_M^{(j)}(P,T),
\label{eq:factorisationDeltaf}
\ea
where it was assumed that the momentum-dependent factor $S/B_M^{(j)}(P,T)$ could also depend on the temperature, but no other thermodynamical quantities. The subscript $X$ was used to identify the spatial part of $S$ and $B$, while the subscript $M$ is used for the momentum dependent term. The sum over $j$ is necessary if e.g. $B(P,X)$ cannot be factorised as $B_S(X) B_M(P)$ but is factorisable as a sum of such terms ($B^{(1)}_S(X) B^{(1)}_M(P)+B^{(2)}_S(X) B^{(2)}_M(P)$). It will be shown shortly that such a general form is appropriate for  the $\delta f_{B/F}(P,X)$ used in this work. More generally, this factorization and the expansion in Eq. (\ref{eq:factorisationDeltaf}) can be seen as an expansion of irreducible tensors in momentum space \cite{Denicol:2012cn}. 

Using Eq.~(\ref{eq:factorisationDeltaf}), the effect of viscosity on photon production (Eq.~(\ref{eq:photonProductionKinVisc})) can be written
\ba
\omega \frac{d^3 R_{\gamma}}{d ^3 k}^{\rm (visc)} &= & \pi_{\mu\nu}(X) K^\mu K^\nu \sum_j S_X^{(j)}(X) \tilde{S}_M^{(j)}(K,T)\nn \\
& &+ \Pi(X) \sum_j B_X^{(j)}(X) \tilde{B}_M^{(j)}(K,T)\;,
\label{eq:photonViscExpansion}
\ea
where $\pi_{\mu\nu}(X) g^{\mu\nu}=0$ and $\pi_{\mu\nu}(X) u^{\mu}=0$  were used again to constrain the coefficient multiplying $\pi_{\mu\nu}(X) $.

The coefficient $\tilde{S}_M^{(j)}(K,T)$ is given by \cite{Shen:2014nfa}
\ba
&&\tilde{S}_M^{(j)}(K,T)  = \frac{1}{2 (K\cdot u)^2}\nn\\
&&\times \left[ g_{\mu\nu}+2u_\mu u_\nu  + 3 \left(\frac{K_\mu K_\nu}{(K\cdot u)^2}-\frac{(K_\mu u_\nu+u_\mu K_\nu)}{(K\cdot u)}\right) \right] \nn \\
 && \quad \times \frac{1}{2(2\pi)^3} \int \frac{d^3 p_1}{2 P^0_1 (2\pi)^3} \frac{d^3 p_2}{2 P^0_2 (2\pi)^3} \frac{d^3 p_3}{2 P^0_3 (2\pi)^3} \nn\\
 &&\quad \times (2\pi)^4\delta^4(P_1+P_2-P_3-K) |\mathcal{M}|^2 \nn \\
& & \quad \times  \left[ \left( P^\mu_1 P^\nu_1 S_M^{(j)}(P_1) \right) f^{(0)}_{B/F}(P_2) \left(1+\sigma_{B/F} f^{(0)}_{B/F}(P_3)\right) \right. \nn \\
& & \left. \qquad  + f^{(0)}_{B/F}(P_1) \left(P^\mu_2 P^\nu_2 S_M^{(j)}(P_2) \right) \left(1+\sigma_{B/F} f^{(0)}_{B/F}(P_3)\right) \right. \nn \\
& & \left. \qquad\qquad  + f^{(0)}_{B/F}(P_1) f^{(0)}_{B/F}(P_2) \sigma_{B/F} P^\mu_3 P^\nu_3 S_M^{(j)}(P_3)  \right], \nn \\
\label{eq:photonProductionKinShearVisc}
\ea
while $\tilde{B}_M^{(j)}(K,T)$ is given by the simpler expression
\ba
&& \tilde{B}_M^{(j)}(K,T)  =  \frac{1}{2(2\pi)^3} \int \frac{d^3 p_1}{2 P^0_1 (2\pi)^3} \frac{d^3 p_2}{2 P^0_2 (2\pi)^3} \frac{d^3 p_3}{2 P^0_3 (2\pi)^3}\nn\\
&&\quad \times  (2\pi)^4\delta^4(P_1+P_2-P_3-K) |\mathcal{M}|^2 \nn \\
& & \quad \times  \left[ \left( B_M^{(j)}(P_1) \right) f^{(0)}_{B/F}(P_2) (1+\sigma_{B/F} f^{(0)}_{B/F}(P_3)) \right. \nn \\
& & \left. \qquad  + f^{(0)}_{B/F}(P_1) \left( B_M^{(j)}(P_2) \right) (1+\sigma_{B/F} f^{(0)}_{B/F}(P_3)) \right. \nn \\
& & \left. \qquad\qquad  + f^{(0)}_{B/F}(P_1) f^{(0)}_{B/F}(P_2) \left(\sigma_{B/F} B_M^{(j)}(P_3) \right) \right]. \nn \\
\label{eq:photonProductionKinBulkVisc}
\ea

Since $\tilde{S}_M^{(j)}(K,T)$ and $\tilde{B}_M^{(j)}(K,T)$ are scalars, they can only depend on $K$ through the combination $K\cdot u$.
It is thus possible to evaluate $\tilde{S}_M$ and $\tilde{B}_M$ in the restframe of the fluid where the photon energy is $\omega = K\cdot u$ and the temperature is $T$.
Although  $\tilde{S}_M^{(j)}(K,T)$ and $\tilde{B}_M^{(j)}(K,T)$ cannot generally be reduced to an analytical expression, it is nevertheless possible to tabulate them as functions of $\omega$ and $T$.

Putting all of this together, the final expression for the photon emission rate is thus\footnote{Note that if photon production rate were to be computed using field-theoretical techniques, then \cite{Kapusta:2006pm}
\ba
\omega \frac{d^3 R_\gamma}{d^3 k} = - \frac{1}{(2 \pi)^3} {\rm Im\, } \Pi^{{\rm R} \mu}_\mu (\omega, \vec k) \frac{1}{\left(e^{\beta \omega} - 1\right)}\;,
\ea
where $ \Pi^{{\rm R} \mu}_\mu (\omega, \vec k)$ is the retarded, finite-temperature, photon self-energy. This equation is exact in the strong interaction, but correct to leading order in the electromagnetic coupling, $\alpha$. 
In that case, linearization in $\delta f_{B/F}(P,X)$ can still be used to write the photon rate as Eq.~(\ref{eq:viscousPhotonRateGeneral}), although with different expressions for $\tilde{S}_M^{(j)}(K)$ and $\tilde{B}_M^{(j)}(K)$. 
}
\ba
\omega \frac{d^3 R_{\gamma}}{d^3 k}&= &\omega \frac{d^3 R_{\gamma}^{(0)}}{d^3 k} \nn\\
&&+\, \pi_{\mu\nu}(X) K^\mu K^\nu \sum_j S_X^{(j)}(X) \tilde{S}_M^{(j)}(K,T) \nn\\
&&+ \,\Pi(X) \sum_j B_X^{(j)}(X) \tilde{B}_M^{(j)}(K,T)\;.
\label{eq:viscousPhotonRateGeneral}
\ea

Using the approximations for the hadron/parton distribution functions outlined in Appendices (\ref{appendixA}) and (\ref{appendixB}), one may now summarize the corrections to the distribution functions that arise from the inclusion of shear and bulk viscosity. For temperatures where the degrees of freedom are partonic, 
\ba
\delta f^{QGP}_{B/F}(P,X) &=& \pi_{\mu\nu}(X) P^\mu P^\nu S_X (X) S_M(P,T) \nn\\
&&+ \Pi(X) B_X^{QGP}(X)  B_M^{QGP}(P,T)\;,
\ea
with (suppressing  arguments)
\ba
S_X&=&\frac{1}{2(\epsilon+\mathcal{P})};  \ \  S_M=\frac{f_{B/F}^{(0)} \left(1 + \sigma_{B/F} f_{B/F}^{(0)} \right)}{T^2} \nn \\
B_X^{QGP}&=&- \frac{1}{15 \left( \frac{1}{3}-c_s^2 \right)\left( \epsilon+\mathcal{P} \right)}   \nn\\
 B_M^{QGP}&=&f_{B/F}^{(0)}(P) \left(1 + \sigma_{B/F} f_{B/F}^{(0)}(P) \right) \nn \\
& & \times \left[ \frac{m^2}{T^2} \frac{T}{P\cdot u}-\frac{P\cdot u}{T} \right]. \nn \\
\ea
For hadronic degrees of freedom (a ``hadronic gas'' [HG]), it is given by
\ba
\delta f^{HG}_{B/F}(P,X) &=& \pi_{\mu\nu}(X) P^\mu P^\nu S_X(X)  S_M(P,T) \nn \\
& & + \Pi(X) \left[ B_X^{HG,1}(X) B_M^{HG,1}(P,T)\right.  \nn\\
&&\left.\qquad + B_X^{HG,2}(X) B_M^{HG,2}(P,T) \right],
\ea
with
\ba
&&B_X^{HG,1}=-\frac{\tau_\Pi}{\zeta} ;\ \ \ B_X^{HG,2}=-\frac{\tau_\Pi}{\zeta}  \left( \frac{1}{3}-c_s^2 \right) \nn\\
&&B_M^{HG,1}=f_{B/F}^{(0)}(P) \left(1 + \sigma_{B/F} f_{B/F}^{(0)}(P) \right) \frac{1}{3} \frac{m^2}{T^2} \frac{T}{P\cdot u}  \nn \\
& & B_M^{HG,2}=f_{B/F}^{(0)}(P) \left(1 + \sigma_{B/F} f_{B/F}^{(0)}(P) \right) \left(-\frac{P\cdot u}{T} \right). \nn \\
\ea
The above decompositions are not uniquely defined, since temperature factors and constants can be in either coefficients. This is not a problem as long as the above definitions are used consistently. These equations thus define the $S$ and $B$ functions that enter the calculation of the viscous photon rates. Then, care must be taken in evaluating  $\tilde{S}_M^{(j)}(K,T)$ and $\tilde{B}_M^{(j)}(K,T)$: a discussion relevant for photon production in the QGP through $2\to 2$ scattering at leading order in the strong coupling constant appears in Ref. (\cite{Shen:2014nfa}).

At present, not all photon sources known are amenable to a calculation of viscous (shear and bulk) corrections. The situation is summarized in Table \ref{table:rates}, together with the appropriate references.

\section{Evaluating the photon momentum anisotropy}

\label{sec:vnTheory}

Before comparisons with measurements are made, it is necessary to explain how photon momentum anisotropies were evaluated in this work, in view of the subtleties involved in their calculation. 

There is a wide literature in heavy ion physics on the different methods of studying the azimuthal anisotropic flow of final state particles (see e.g. \cite{Luzum:2013yya} for recent review). The azimuthal dependence of a given particle's underlying momentum distribution, in a particular event, is usually characterized by a Fourier series written in the form of a Fourier coefficient $v^s_n$ and event-plane angle $\Psi^s_n$:
\be
v^s_n e^{i n \Psi^s_n} = \frac{\int d p_T  dy d\phi p_T \left[ p^0 \frac{d^3 N^s}{d^3 p} \right] e^{i n\phi}}{\int d p_T dy  d\phi p_T \left[ p^0 \frac{d^3 N^s}{d^3 p} \right]} \;.
\label{eq:vnOneEv}
\ee
The superscript ``$s$'' denotes the particle species corresponding to the underlying momentum distribution. It is possible to assign additional labels identifying the kinematic cuts used on $p_T$ and $y$, but this is not necessary in what follows, since there should not be any ambiguity about the cuts used for each particle species.

Experiments measure samples of the distribution $p^0 d^3 N^s/d^3 p$, averaged over numerous heavy ion collisions. Given an appropriate measurement of azimuthal correlation between different particles, these event-averaged measurements can be mapped to event-averages of $v_n$ and $\Psi_n$. These can in turn be computed from theoretical models which can often access the full $p^0 d^3 N^s/d^3 p$ distribution.

Due to the limited statistics available, photon anisotropy measurements are not photon-photon correlations, but rather photon-hadron correlations. Both experiments that have measured photon anisotropies, PHENIX at RHIC and ALICE at the LHC, used the event-plane method~\cite{Poskanzer:1998yz} to make the measurement. This method can be understood as using hadrons to define an effective reference plane in the transverse direction, based on the hadron's azimuthal distribution; the photon momentum anisotropy is then measured with respect to this hadronic plane. Depending on the number of hadrons being measured and on the size of their azimuthal momentum anisotropy, the hadronic event-plane cannot necessarily be reconstructed accurately. This introduces a small uncertainty, of the order of $10\%$, in the mapping of event-plane method measurement to the $v_n$ and $\Psi_n$ of photons and hadrons~\cite{Alver:2008zza,Ollitrault:2009ie,Luzum:2012da}.

The two limits of event-plane method measurements are known as the low and high resolution limits. In the low resolution limit, the event-plane anisotropy reduces to the scalar product $v_n\{SP\}$ anisotropy~\cite{Luzum:2013yya}:
\be
v_n\{EP\} \overset{\textrm{low res.}}{=} v_n\{SP\}=\frac{\langle v_n^{\gamma} v_n^{h} \cos(n(\Psi_n^{\gamma}-\Psi_n^{h})) \rangle}{\sqrt{\langle ( v_n^{h} )^2 \rangle}} \;. \nn \\
\label{eq:v2SP}
\ee
The other limit is the high resolution limit: 
\be
v_n\{EP\} \overset{\textrm{high res.}}{=} \langle v^\gamma_n \cos(n(\Psi_n^\gamma-\Psi_n^h)) \rangle .
\label{eq:vnAverage}
\ee
The angle brackets $\langle \ldots \rangle$ represents an average over events.

The resolution correction, which quantifies the accuracy of the event-plane reconstruction, is used to determine which limit of $v_n\{EP\}$ should be compared with measurements. Their value for both RHIC and LHC measurements are shown respectively in Refs.~\cite{Adare:2015lcd} and \cite{LohnerThesis}. The value of the resolution correction changes with the centrality and methods used to determine the event-plane. For $n=2$ ($v_2\{EP\}$), it is neither clearly in the high nor low resolution limit. On the other hand, higher harmonics ($n>2$) are closer to the low resolution limit. Equation~(\ref{eq:v2SP}) is thus used in this work to evaluate $v_n\{EP\}$. It was verified that the other limit of $v_n\{EP\}$, Eq.~(\ref{eq:vnAverage}), does not differ from Eq.~(\ref{eq:v2SP}) by more than 10\%~\cite{Paquet:2015Thesis}. The uncertainty associated with this ambiguity in $v_n\{EP\}$ is thus not a significant issue.

The experimental measurements correlate hadrons from a wide bin in $p_T$ to photons measured in a small $p_T$ bin, effectively resulting in a $v^\gamma_n\{EP\}$ differential in the photon transverse momentum:
\be
v^\gamma_n\{EP\}(p^\gamma_T)\overset{\textrm{low res.}}{=}\frac{\langle v_n^{\gamma}(p^\gamma_T) v_n^{h} \cos(n(\Psi_n^{\gamma}(p^\gamma_T)-\Psi_n^{h})) \rangle}{\sqrt{\langle ( v_n^{h} )^2 \rangle}}, \nn \\
\label{eq:v2SPdiff}
\ee
where the $h$ superscript refers to the charged hadrons with which photons are correlated. In this work we evaluate $v_n^h$ and $\Psi_n^h$ at midrapidity, integrated over $p_T>0.3$~GeV. It was verified that the result of Eq.~(\ref{eq:v2SPdiff}) did not change with other choices of lower $p_T$-cuts between $0$ and $0.5$~GeV.

Equation~(\ref{eq:v2SPdiff}) assumes that the events that are averaged over have small multiplicity fluctuations. That is, all events are assumed to produce a similar number of photons and hadrons. If large multiplicity fluctuations are present, Eq.~(\ref{eq:v2SPdiff}) will take a different form depending of the details of the measurement, for example whether all events are treated equally, or if events with more hadrons and photons are given a larger weight in the event-average.

To reduce the importance of multiplicity fluctuations, experimental collaborations first measure $v^\gamma_n\{EP\}$ in small centrality bins~\cite{LohnerThesis}. The anisotropy measurements from these smaller centrality bins are then recombined into a larger centrality to reduce the statistical uncertainty of the measurement. When the small centrality bins are recombined, each centrality is weighted by the number of photons measured in the centrality~\cite{LohnerThesis}:
\be
v^\gamma_n\{EP\}[c_{\textrm{min}},c_{\textrm{max}}]=\frac{\sum_{c \in [c_{\textrm{min}},c_{\textrm{max}}]} v^\gamma_n\{EP\}[c] N[c]}{\sum_{c \in [c_{\textrm{min}},c_{\textrm{max}}]} N[c]}, \nn \\
\label{eq:vnCent}
\ee
where $N[c]$ is the number of photons measured in centrality $c$, $v^\gamma_n\{EP\}[c]$ is the momentum anisotropy measured in $c$ and $[c_{\textrm{min}},c_{\textrm{max}}]$ is the final (large) centrality class in which the measurement is reported. At the LHC the sub-bins are~\cite{LohnerThesis} $0-5\%$, $5-10\%$, $10-20\%$, $20-30\%$ and $30-40\%$, while $10\%$ bins are used at RHIC~\cite{BannierThesis}.

The quantity $v^\gamma_n\{EP\}[c_{\textrm{min}},c_{\textrm{max}}]$ --- Eq.~(\ref{eq:vnCent}) --- is the one that should be compared to PHENIX and ALICE measurements. All photon anisotropy calculations presented in this paper are computed with Eq.~(\ref{eq:vnCent}) using the bins just listed.

\section{Results and discussion}

We now show and discuss the result of integrating the photon rates discussed in Sections~\ref{sec:thermal} and \ref{sec:visc_rates}, with the hydrodynamic approach discussed in Section \ref{sec:hydro}. Prior to doing this, an important clarification is needed. The model used here is a hybrid approach, in the sense that it is not purely hydrodynamics: it has a viscous fluid-dynamics stage that is followed by a transport phase -- modelled with UrQMD -- with dynamic decoupling. The UrQMD afterburner is important to a successful theoretical interpretation of the measured proton spectra and $v_2$ \cite{Ryu:2015vwa,Ryu_prep}. 
However, extracting the photons via the vector meson spectral density~\cite{Turbide:2003si} from a transport model is still very much a topical subject of current research.
More generally, electromagnetic emissivities are typically calculated in conditions near thermal equilibrium, as discussed earlier in this paper, and a knowledge of the local temperature and of other thermodynamic variables is usually absent from most transport formulations. One resolution of this situation has been to coarse-grain the transport final states, and to assign local temperatures to cells on a space-time grid using the equation of state \cite{Huovinen:2002im,Endres:2015fna}. Such procedures are numerically-intensive, but will be studied within our framework in detail in the future. The point of view adopted in this work is that, apart from proton observables,  hydrodynamics does provide a realistic environment for the bulk of hadronic observables, especially if the bulk viscosity is included \cite{Ryu:2015vwa}. Therefore, for the calculation of photons, the contribution of the UrQMD phase of the spatiotemporal evolution is modelled by letting the fluid-dynamical evolution proceed past the switching temperature from hydro to UrQMD (the ``particlization temperature'' \cite{Huovinen:2012is}), $T_{\rm switch} = 145$ MeV, down to a more typical hydro freeze-out temperature of $T = 105$ MeV. In hydrodynamical approaches in general, the freeze-out temperature is a free parameter of the model: more words about the dependence of the photon signal on this parameter will appear later in this section.

\subsection{RHIC}

\begin{figure}[tb]
        \includegraphics[width=0.32\textwidth]{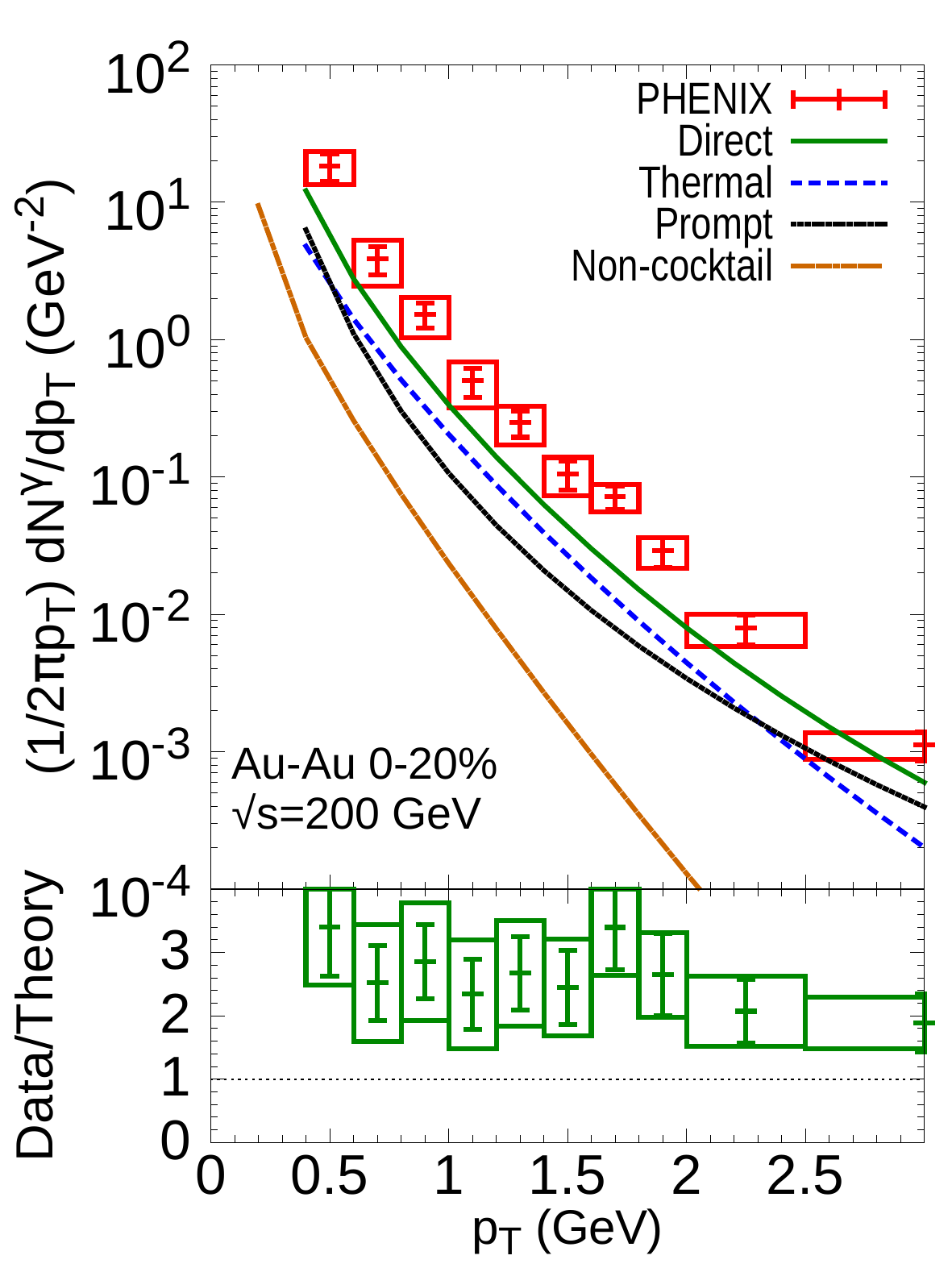}
        \includegraphics[width=0.32\textwidth]{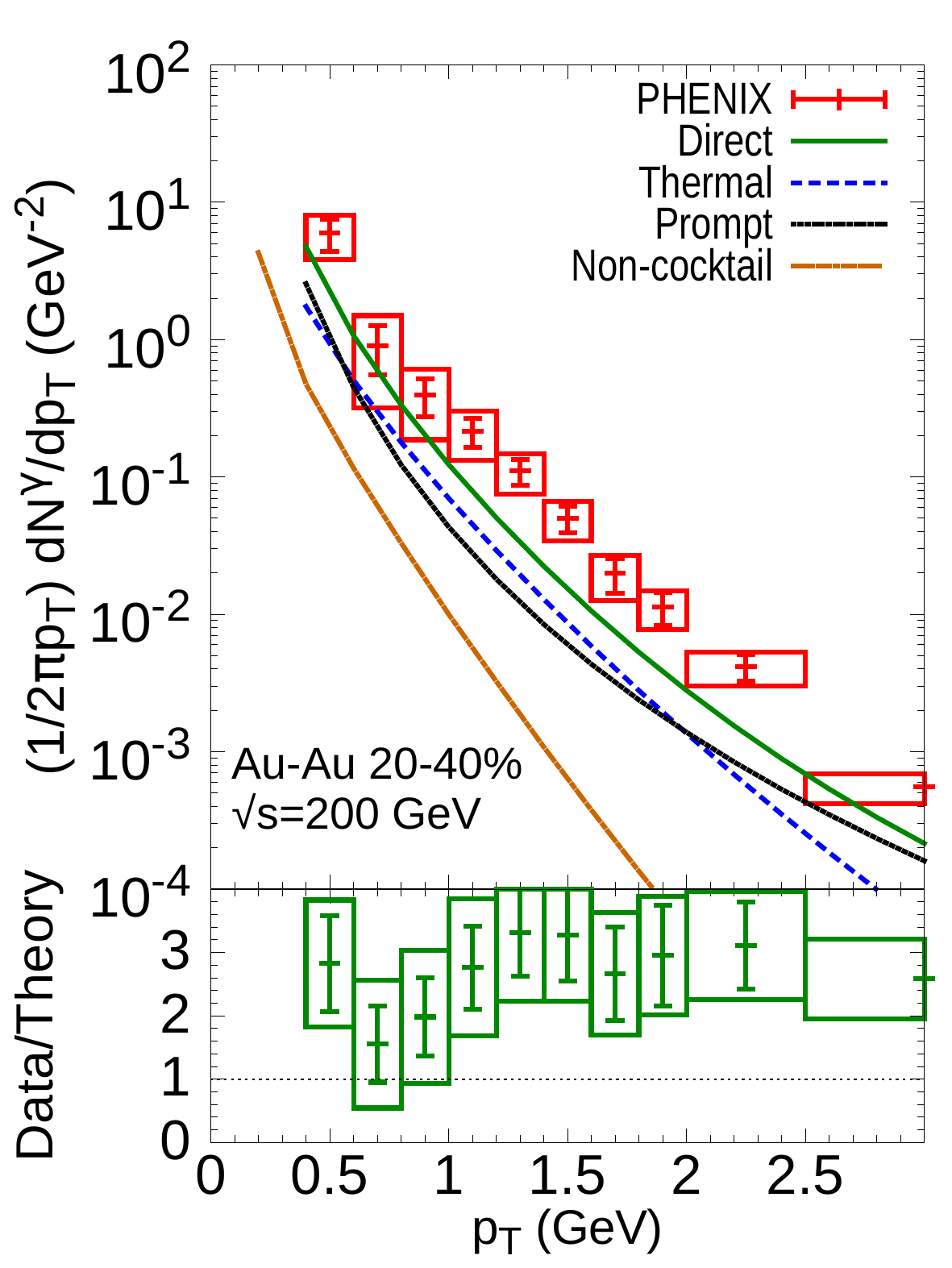}
\caption{The result of a hydrodynamic calculation of direct photon spectra, for Au - Au collisions at RHIC, in the 0 - 20 \% (top panel) and 20 - 40\% (bottom panel) centrality range. The different curves are explained in the text, the data are from Ref.~\cite{Adare:2014fwh}. 
}
\label{fig:directPhotonRhic_spect}        
\end{figure}

\begin{figure}[tb]
				\includegraphics[width=0.32\textwidth]{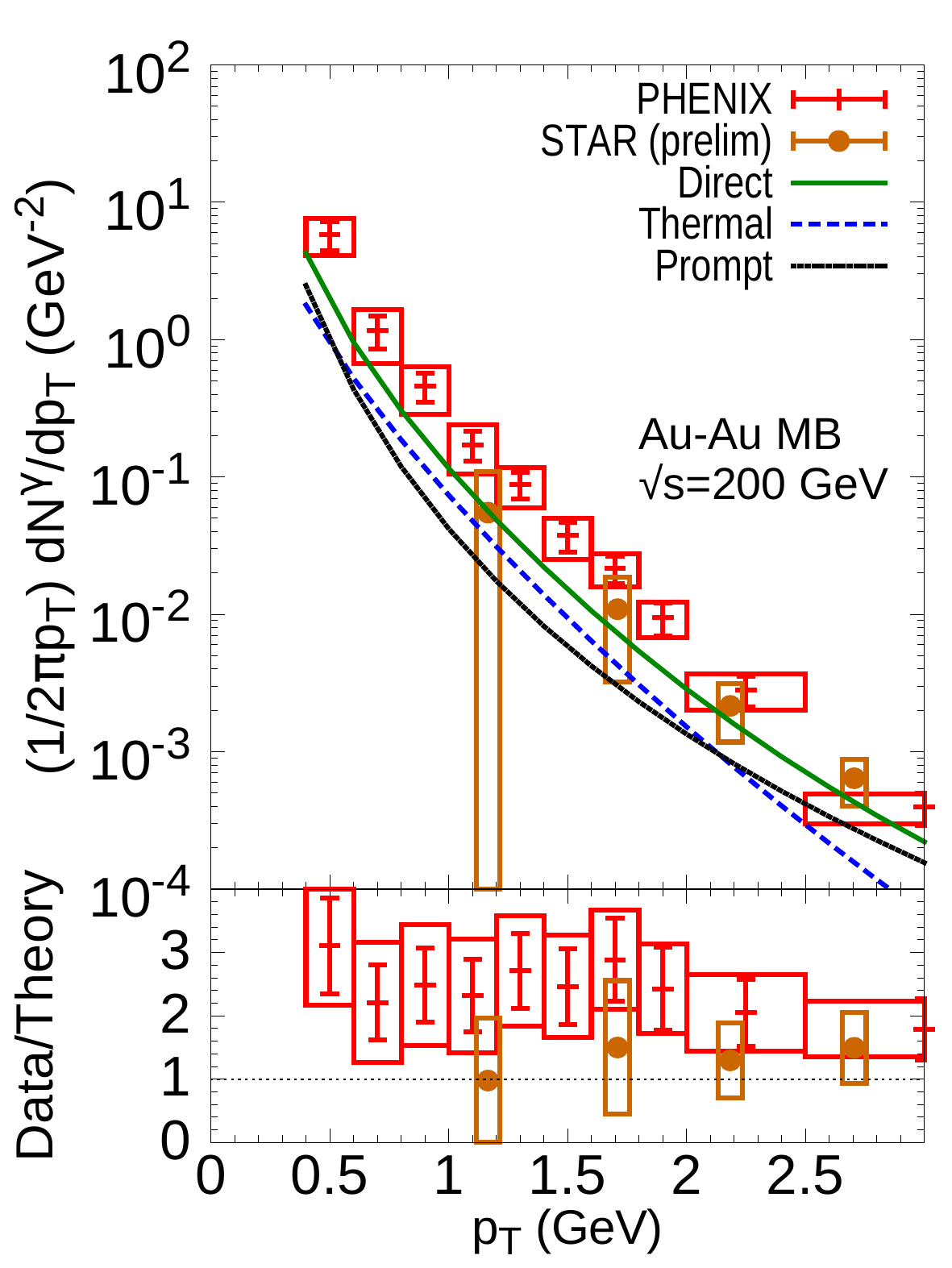}
\caption{The result of a hydrodynamic calculation of direct photon spectra, for Au - Au collisions at RHIC, in minimum bias centrality range. The data are from Refs. \cite{Adare:2014fwh,ChiYangThesis}. 
}
\label{fig:directPhotonRhic_spect_MB}        
\end{figure}

The direct photon spectrum and $v_2$ were  measured at RHIC by the PHENIX collaboration~\cite{Adare:2014fwh,Adare:2011zr,Bannier:2014bja,Adare:2015lcd}. These measurements were made in Au-Au collisions at $\sqrt{s_{NN}}=200$~GeV for centralities 0-20\% and 20-40\%. Comparison of the hydrodynamical model's results for direct photon spectra are shown in Fig. \ref{fig:directPhotonRhic_spect}.
The preliminary, minimum-bias direct photon spectrum measurement from STAR~\cite{ChiYangThesis} is shown in Fig.~\ref{fig:directPhotonRhic_spect_MB}, and is compared with both the hydrodynamical calculations and the PHENIX measurements from Ref.~\cite{Adare:2014fwh}.
The dashed lines represent the thermal contributions, that is the sum of all contributions of thermal origin. The prompt photons are calculated in NLO QCD, as explained earlier. The contribution of non-cocktail photons (Section~\ref{sec:nonCocktail}) is also shown.

\begin{figure}[tb]
       \includegraphics[width=0.35\textwidth]{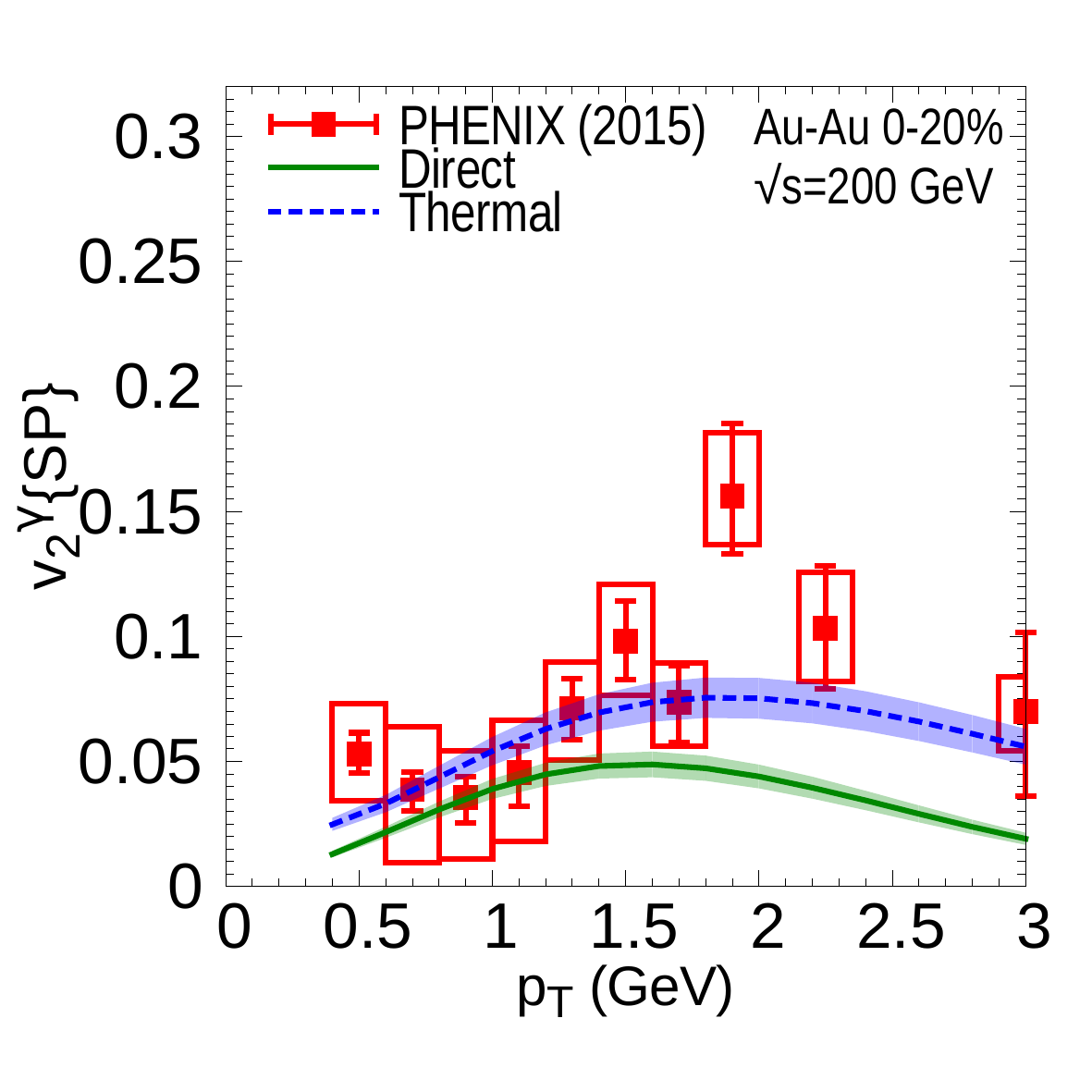}
       \includegraphics[width=0.35\textwidth]{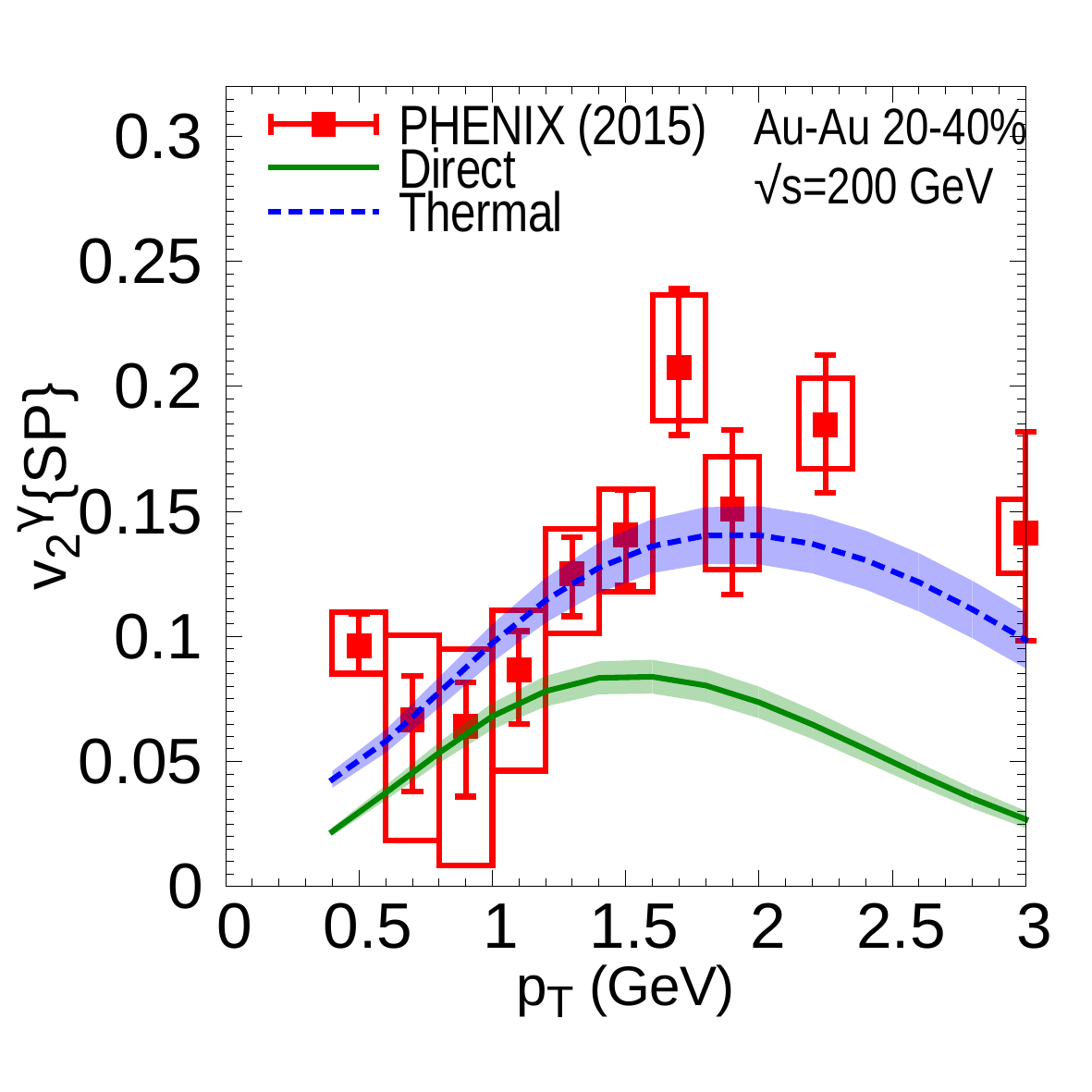}
\caption{Hydrodynamic calculation of the direct photon $v_2$, for Au - Au collisions at RHIC, in the 0 - 20 \% (top panel) and 20 - 40\% (bottom panel) centrality range . The data are from Ref.~\cite{Adare:2015lcd}.} 
\label{fig:directPhotonRhic_v2}
\end{figure}

The curves labeled ``direct'' represent the sum of all sources considered in this work (Section~\ref{sec:photonSources}). One observes that the calculation, with the contributions enumerated in the text, and the experimental tend to converge for values of $p_T \gtrsim 2.5$ GeV. There, the calculation almost entirely consists of the pQCD component. For intermediate transverse momenta (as defined by this figure, $p_T \approx 1.5$ GeV), the calculation underestimates the PHENIX data central points roughly by a factor of~3. 
Agreement of the calculations with the preliminary STAR data (Fig.~\ref{fig:directPhotonRhic_spect_MB}) is considerably better, well within systematic uncertainties.

In the low $p_T$ region, calculation and data are reunited again, but bear in mind the strong caveats regarding the trustworthiness of the pQCD calculations at such low transverse momenta. As supported by a direct comparison with pp photon data, the prompt photon curve shown in Figs. \ref{fig:directPhotonRhic_spect} and \ref{fig:directPhotonRhic_spect_MB} should hold down to $p_T \approx$ 1 GeV. While one does not expect a sudden breakdown of the formalism used here, it does becomes less predictive as the photon momentum goes down. The theoretical interpretations of photon production in nucleus-nucleus collisions would rest on much firmer ground if a fundamental measurement of soft photons from pp collisions, extending to values of transverse momenta compared to those in Figs. \ref{fig:directPhotonRhic_spect} and \ref{fig:directPhotonRhic_spect_MB} existed.
Such a measurement, while challenging, would provide a valuable baseline for phenomenological modelling, and would further our understanding of QCD in its strongly coupled regime. 

Figure \ref{fig:directPhotonRhic_v2} shows the calculated photon elliptic flow, compared with data measured by the PHENIX collaboration. The photon anisotropy was evaluated with Eq.~(\ref{eq:vnCent}). The elliptic flow shows the now characteristic shape, with the turnover at $p_T \gtrsim$ 2 GeV driven by the pQCD photons. As was the case for the photon spectra the calculation of the photon elliptic flow systematically undershoots the central data points. However, and this also holds for the spectra, taking into account the statistical and systematic uncertainties greatly reduces the tension between theory and experiment. Thermal photons, represented by the dashed curves, are shown separately to highlight that the thermal contribution does exhibit a large $v_2$, but that this momentum anisotropy is then suppressed by prompt photons. 

As can be expected from their small contribution to the direct photon spectra (Fig.~\ref{fig:directPhotonRhic_spect}), non-cocktail photons do not contribute significantly to the direct $v_2$. They are not shown in Figure~\ref{fig:directPhotonRhic_v2}.

\subsection{LHC}

\begin{figure}[tbh]
        \includegraphics[width=0.33\textwidth]{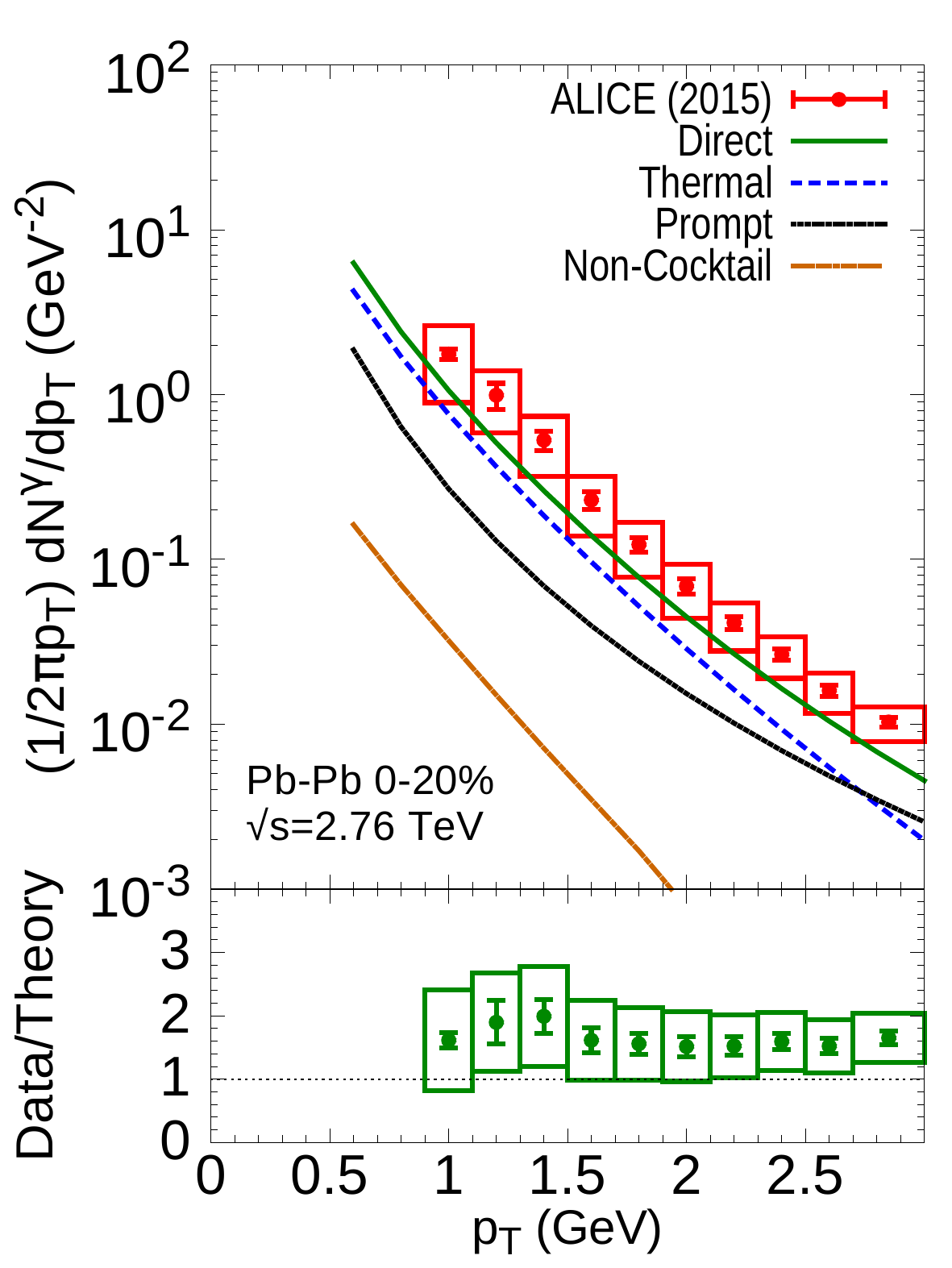}
				\includegraphics[width=0.33\textwidth]{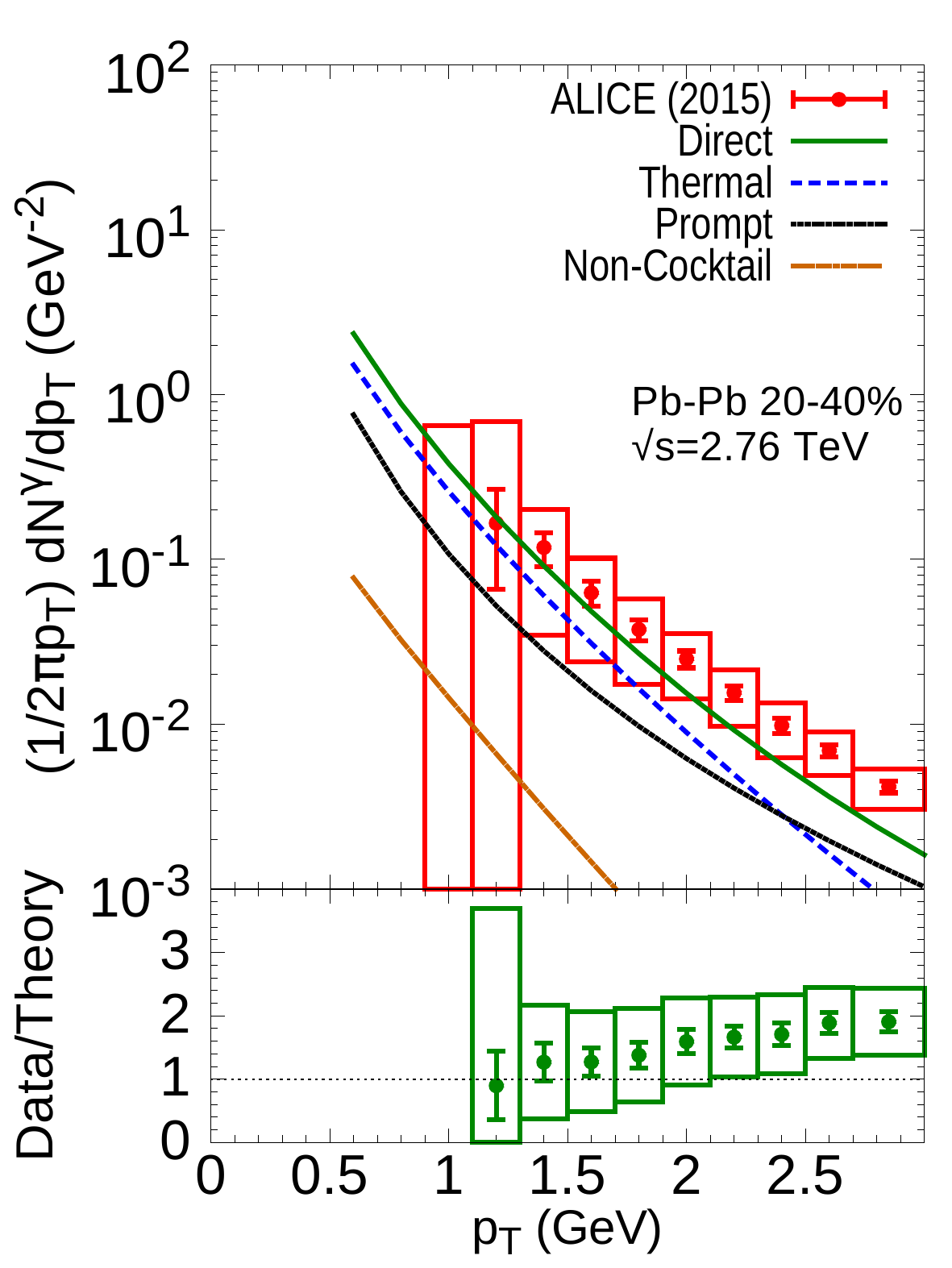}
\caption{The direct photon spectrum for Pb-Pb collisions at the LHC 0 - 20\% (top panel) and 20 - 40\% (bottom panel) centrality range. The different curves are explained in the text, and the data are from the ALICE Collaboration~\cite{Adam:2015lda}.}
\label{fig:directPhotonLHC}        
\end{figure}

\begin{figure}[tbh]
        \includegraphics[width=0.33\textwidth]{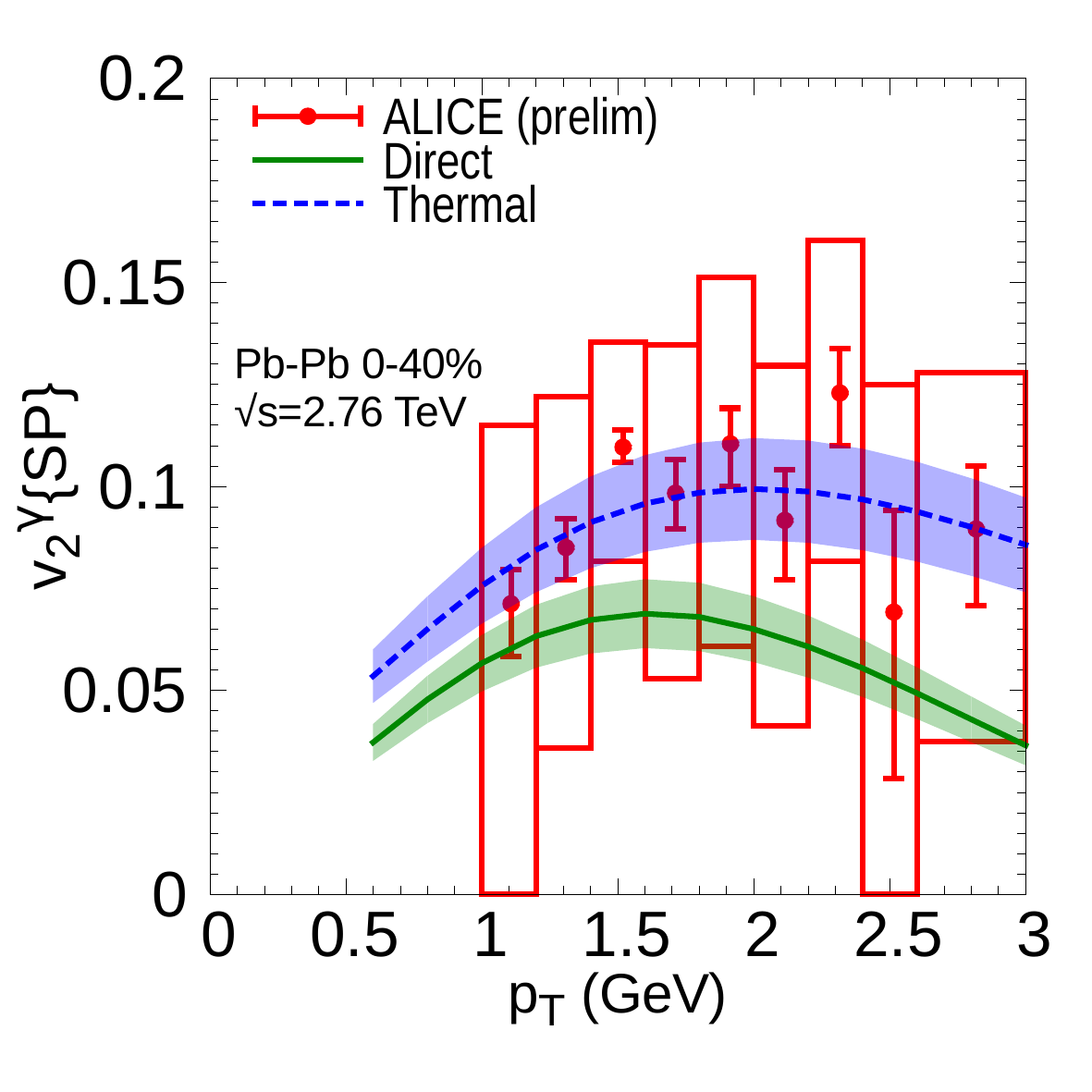}
\caption{The direct photon $v_2$ at 0 - 40\% centrality. Data are from the ALICE Collaboration~\cite{Lohner:2012ct,LohnerThesis}.}
\label{fig:directPhotonV2LHC}        
\end{figure}

The direct photon spectrum and $v_2$ in Pb-Pb collisions at $\sqrt{s_{NN}}=2760$~GeV are presented in Figs.~\ref{fig:directPhotonLHC} and \ref{fig:directPhotonV2LHC} respectively. The calculations are compared with measurements from the ALICE collaboration~\cite{Adam:2015lda,Lohner:2012ct,LohnerThesis}. As for RHIC, the contribution to the spectrum of  prompt, thermal and non-cocktail photons, along with their sum (direct photons), are shown separately in Fig.~\ref{fig:directPhotonLHC}. The elliptic flow of thermal photons and of the total number of direct photons is plotted in Fig.~\ref{fig:directPhotonV2LHC}. The general features of the photon data set at the LHC is reminiscent of that at RHIC, but important differences emerge when comparing with theoretical calculations. For both LHC observables --- spectrum and $v_2$ ---  there is less tension between data and theory than at RHIC. In fact, the theory results are in agreement with the experimental results when considering the statistical and systematic uncertainties. 
As previously, the  prompt contribution begins to take over at around $p_T \approx 3$ GeV, but otherwise lies systematically below the thermal sources.

\subsection{Effect of bulk viscosity}

The calculation of direct photons presented in this work is the first one to include the effect of bulk viscosity on both the medium evolution and the photon emission rates. Considering that the introduction of bulk viscosity was shown to have a large effect on the description of the hadronic observables~\cite{Ryu:2015vwa}, it is important to highlight its effect on photon production.

\begin{figure}[tb]
        \includegraphics[width=0.35\textwidth]{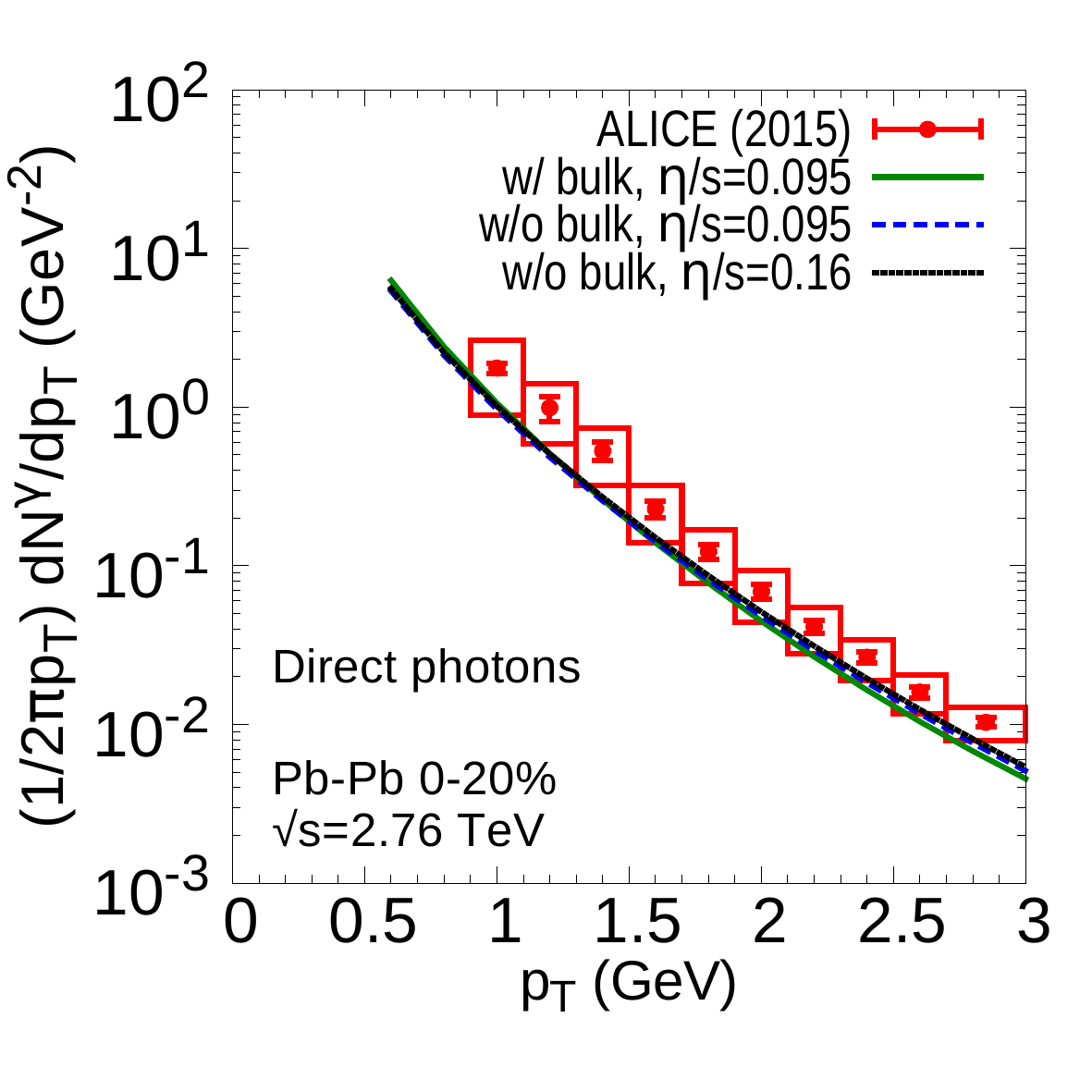}
        \includegraphics[width=0.35\textwidth]{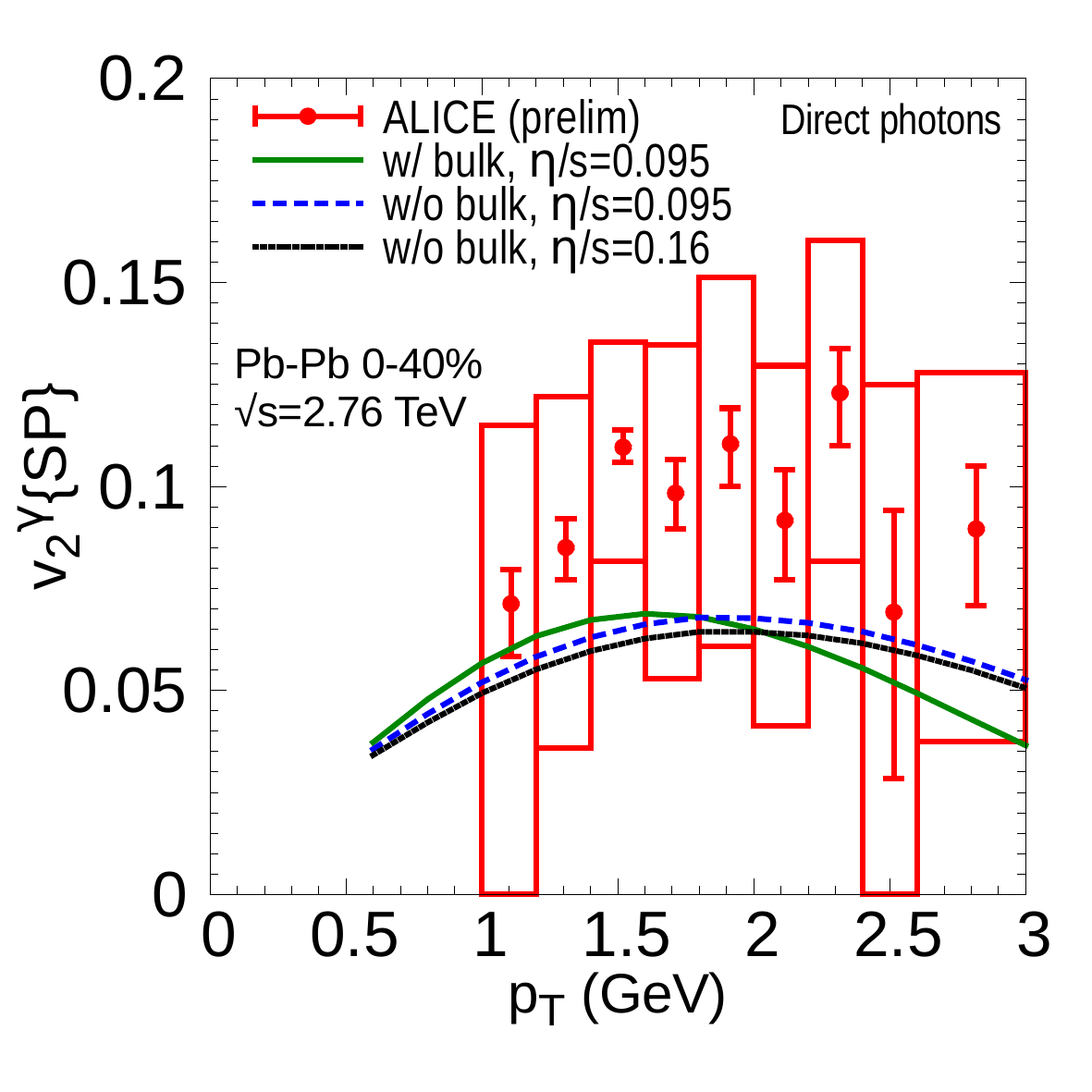}
\caption{Effect of bulk viscosity on the direct photon spectrum (top panel) and $v_2$ (bottom panel) in Pb-Pb collisions at $\sqrt{s_{NN}}=2760$~GeV. ALICE measurements~\cite{Wilde:2012wc,Lohner:2012ct,LohnerThesis} are shown for reference.}
\label{fig:directPhotonBulkVsShear}        
\end{figure}

In Fig.~\ref{fig:directPhotonBulkVsShear}, the direct photon spectrum and $v_2$ are shown with and without bulk viscosity for $\sqrt{s_{NN}}=2760$~GeV Pb-Pb collisions at the LHC. Since the inclusion of bulk viscosity modifies the \emph{shear} viscosity necessary to describe the hadronic momentum anisotropies~\cite{Ryu:2015vwa}, two direct photon calculations without bulk viscosity are shown: one with $\eta/s=0.095$, which is the shear viscosity necessary to describe the hadronic $v_n$ in the presence of bulk viscosity~\cite{Ryu:2015vwa}, and one with $\eta/s=0.16$, for which a good description of hadronic $v_n$ can be achieved with $\zeta/s=0$. It can be seen that the two calculations that do not include bulk viscosity are similar, both for the spectrum and the $v_2$.

The effect of bulk viscosity on the spectrum of direct photons is small, and consists of a slight softening of the spectrum. Like the spectrum, the $v_2$ increases at low $p_T$ and decreases at high $p_T$. This changes the shape of $v_2$, whose maximum value is shifted toward lower $p_T$. This is a distinctive photonic signal of the finite bulk viscosity of QCD around the transition region.

\begin{figure}[tb]
        \includegraphics[width=0.35\textwidth]{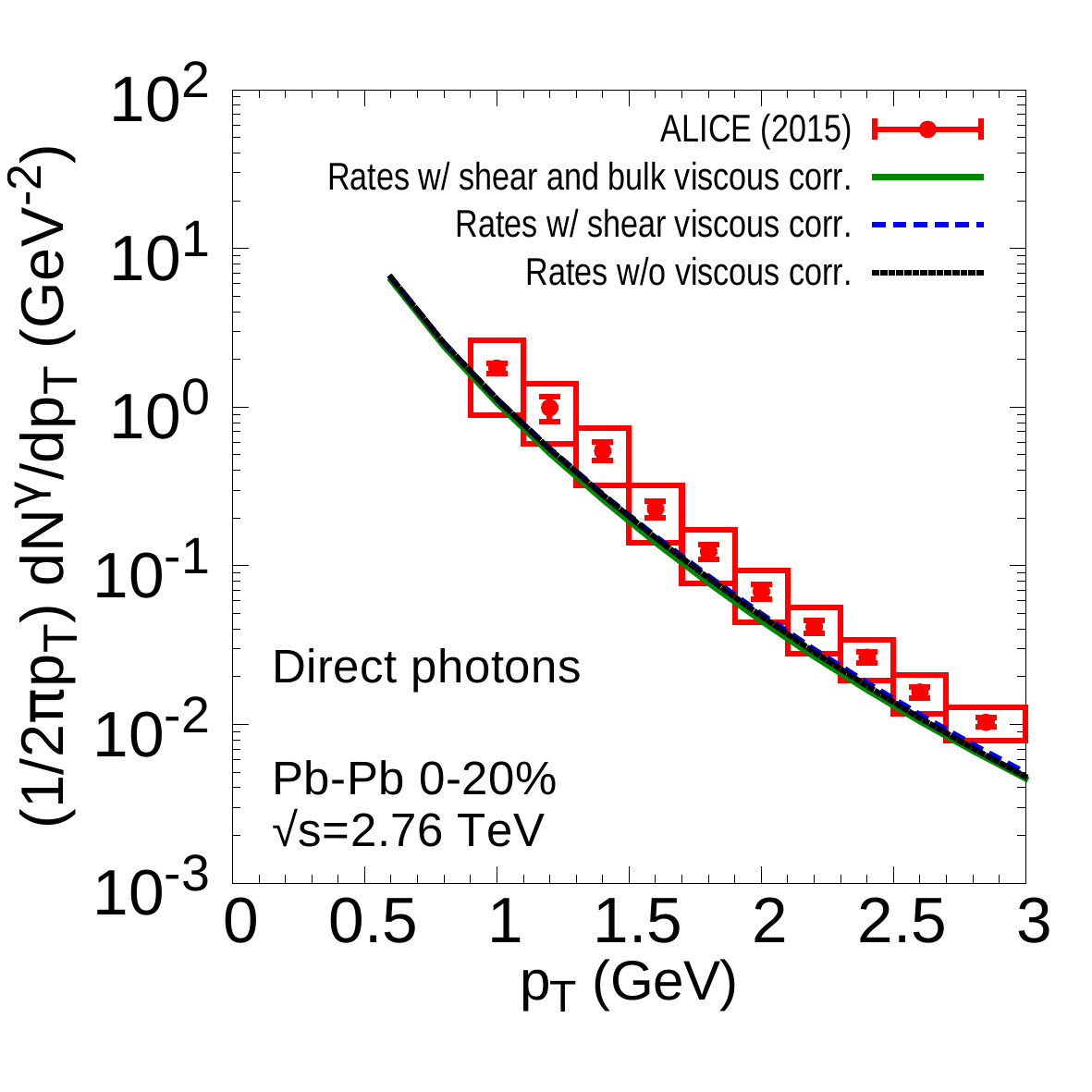}
				\hspace{1cm}
        \includegraphics[width=0.35\textwidth]{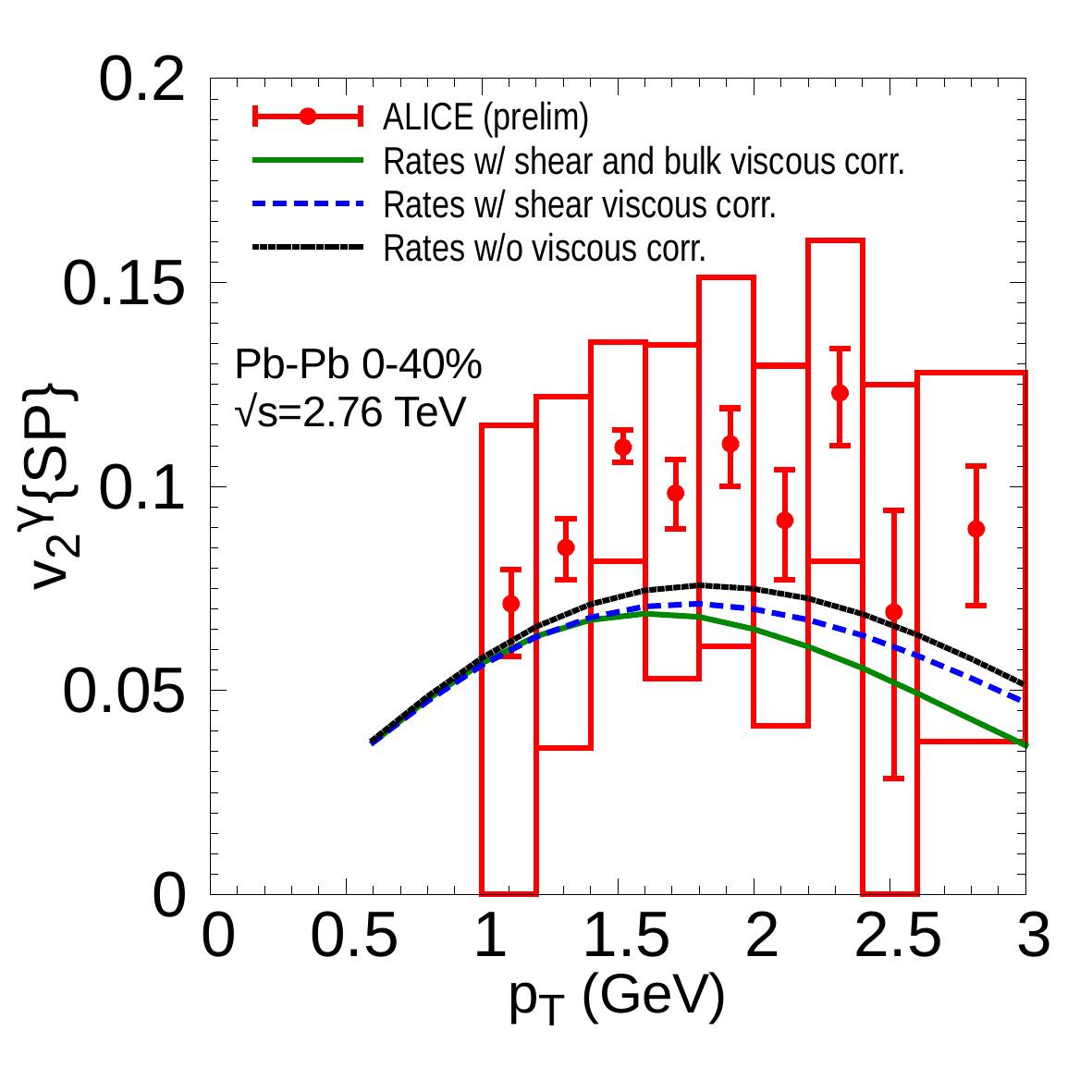}
\caption{Effect of viscosity corrections to the photon emission rates for the direct photon spectrum (top panel) and $v_2$ (bottom panel) in Pb-Pb collisions at $\sqrt{s_{NN}}=2760$~GeV.}
\label{fig:directPhotonRateCorr}        
\end{figure}

The effect of bulk viscosity on direct photons can be divided in two separate contributions: its effect on the photon emission rates (Section~\ref{sec:visc_rates}), and its effect on the spacetime evolution of the medium. The effect of bulk viscosity on the emission rates is illustrated in Fig.~\ref{fig:directPhotonRateCorr} by showing the photon spectrum and $v_2$ with and without corrections to the rates due to bulk viscosity. The effect of the shear viscous correction to the photon rates is shown as well, for reference. 

Viscous corrections to the rates have a small effect on the direct photon spectrum. This can be understood from the fact that viscous corrections are larger at higher $p_T$, where prompt photons dominate over thermal ones. 

The direct photon $v_2$, on the other hand, is suppressed at higher $p_T$ by both shear and bulk viscosity corrections to the photon rates. The suppression is  of the order of $20-30\%$. Recall however that not all photon emission rates are corrected for the effect of shear and bulk viscosities, as listed in Table~\ref{table:rates}. In consequence, the results shown in Fig.~\ref{fig:directPhotonRateCorr} most likely underestimate the effect of viscosity on the photon rates.

\begin{figure}[tbp]
        \includegraphics[width=0.23\textwidth]{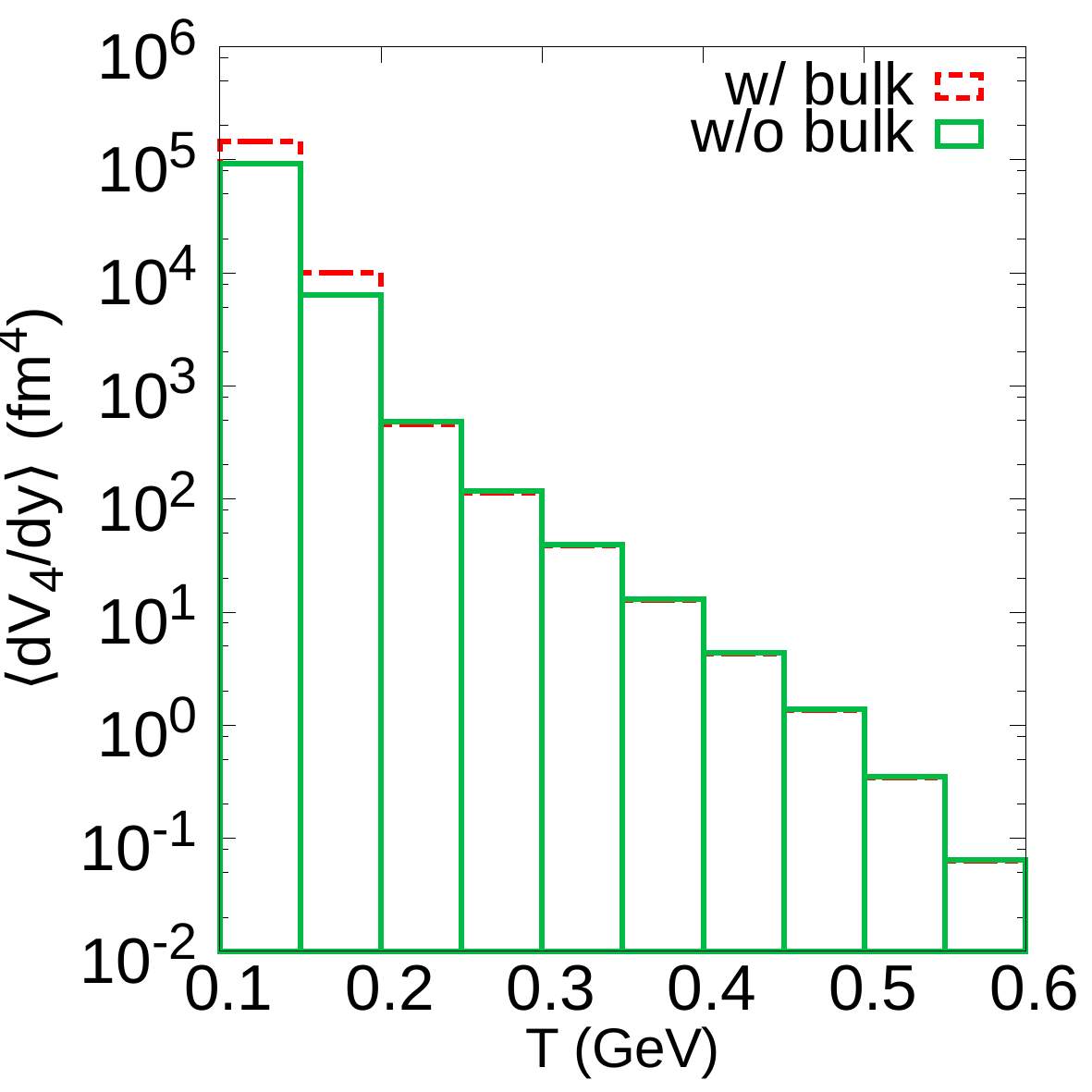}
        \includegraphics[width=0.23\textwidth]{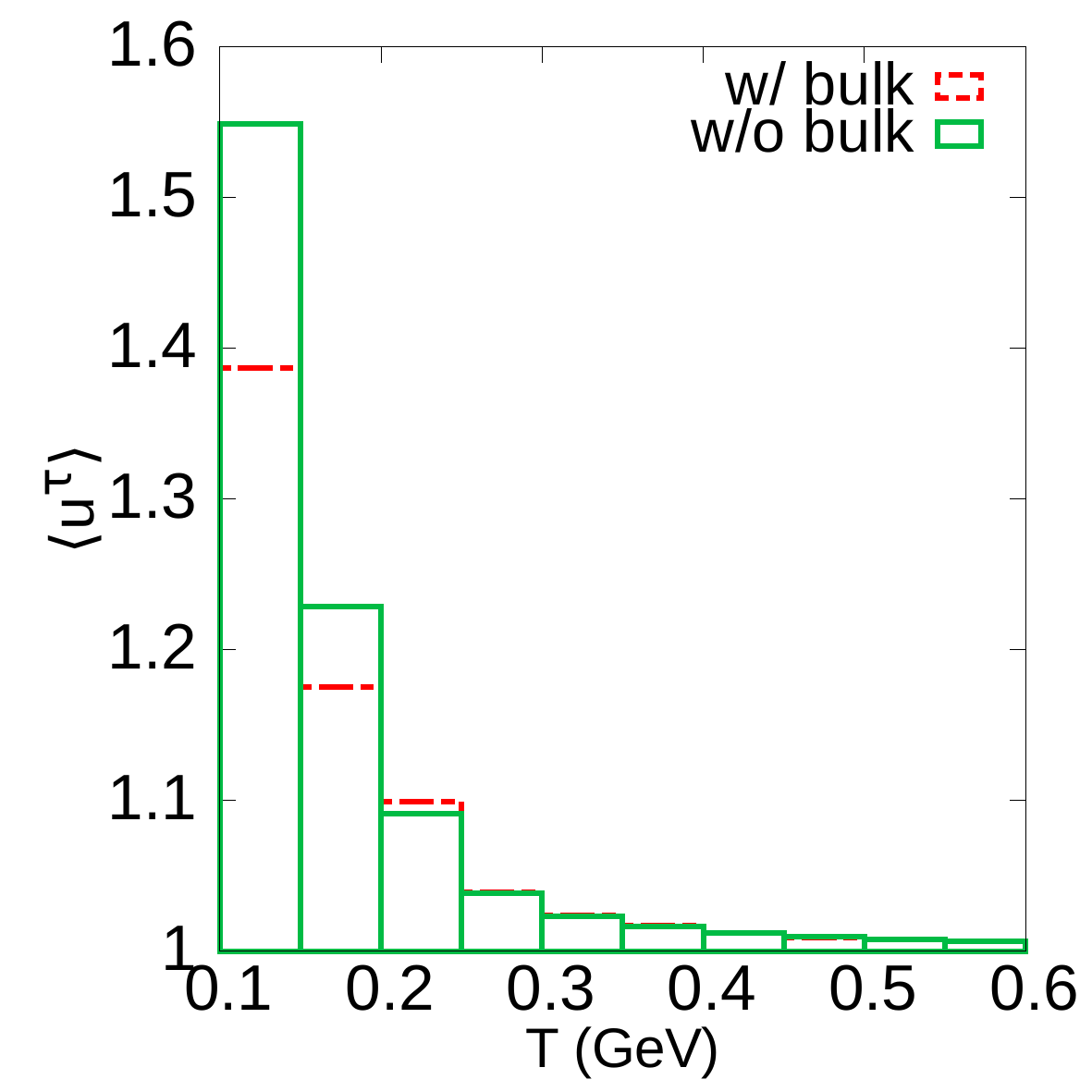}
\caption{Event-average spacetime volume $\left\langle d V_4/d y \right\rangle_T$ (left) and event-average flow velocity $\langle u^\tau \rangle_T$ (right) for hydrodynamical model with and without bulk viscosity in Pb-Pb collisions at $\sqrt{s_{NN}}=2760$~GeV.}

\label{fig:directPhotonSpacetime}        
\end{figure}

The effect of bulk viscosity on the spacetime description of the medium is illustrated in Fig.~\ref{fig:directPhotonSpacetime}. The change in spacetime volume induced by the inclusion of bulk viscosity is shown for different ranges of temperature on the left, while the effect of bulk viscosity on the flow velocity distribution, as quantified by $u^\tau=\sqrt{1+(u^x)^2+(u^y)^2}$, is shown on the right. The effect of bulk viscosity is clear: it reduces the transverse expansion of the medium at low temperature, but considerably increases its spacetime volume. Since thermal photon emission is proportional to the spacetime volume, the increase in volume translates into a larger number of emitted photons. On the other hand the slower transverse expansion implies a softer photon spectrum, with more soft photons emitted but less hard ones. It is the combination of these two effects that produce an overall softening of the photon spectrum in the presence of bulk viscosity.

\subsection{Effect of photon emission rates}

\begin{figure}[tb]
        \includegraphics[width=0.35\textwidth]{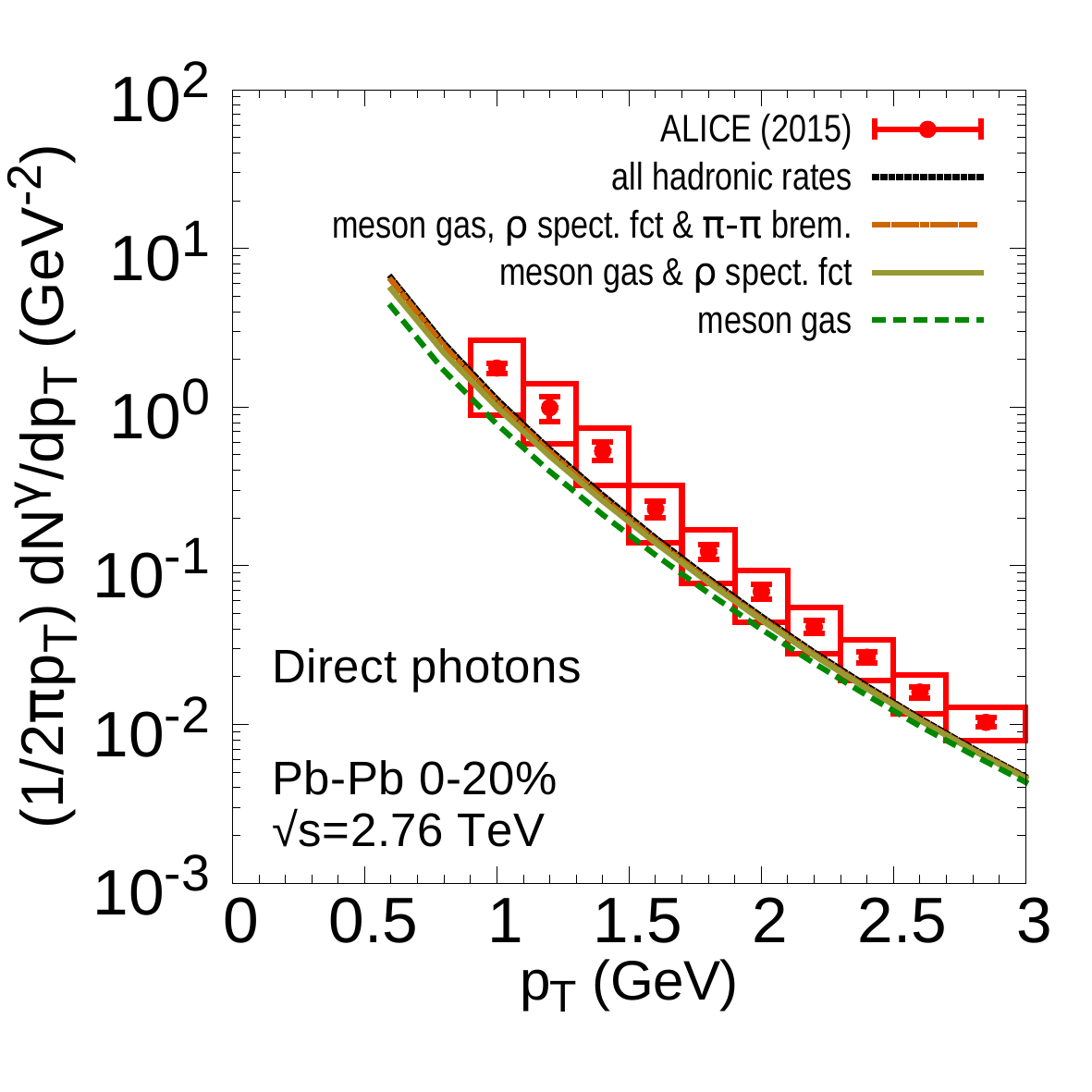}
        \includegraphics[width=0.35\textwidth]{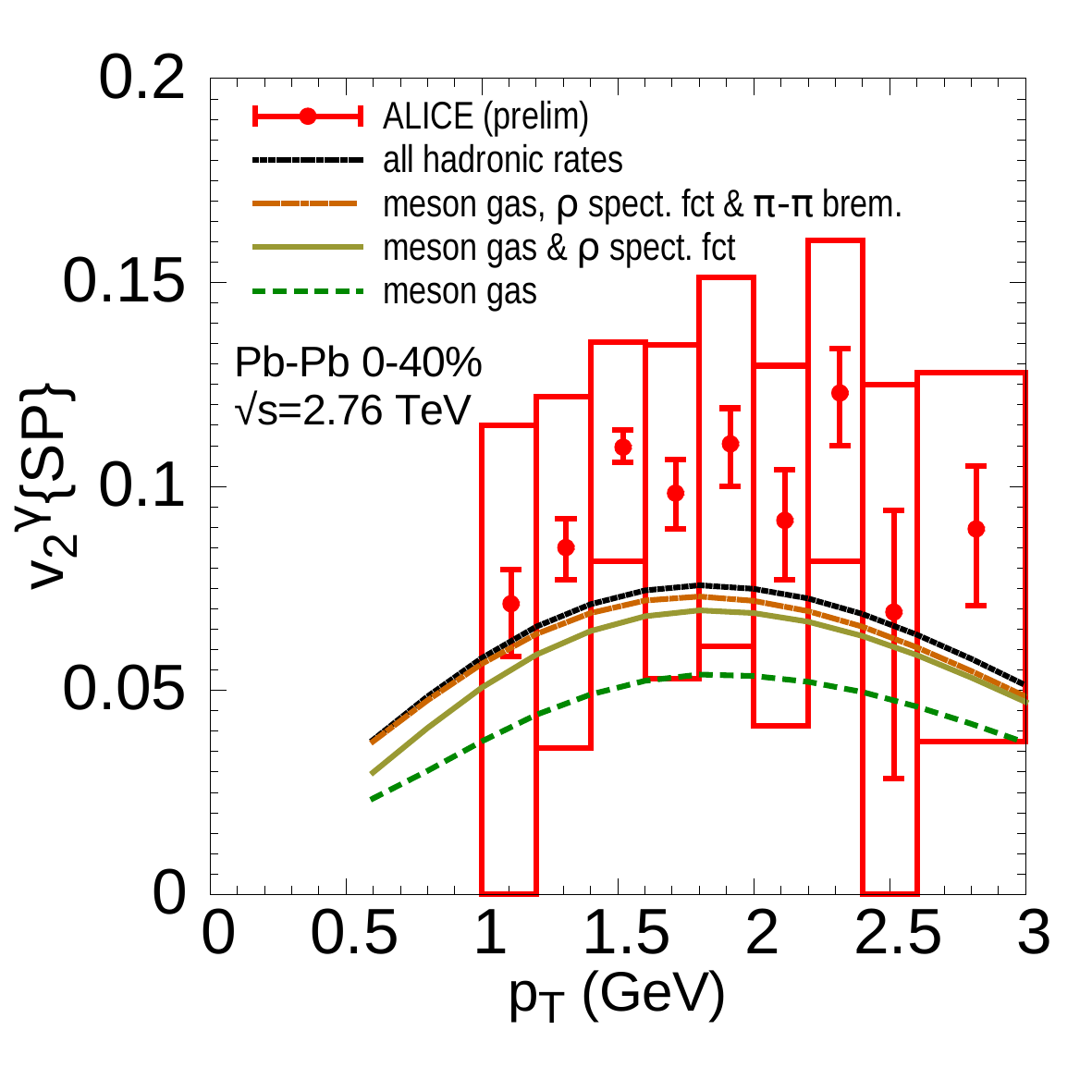}
\caption{Importance of different hadronic photon production channels on the direct photon spectrum (top) and $v_2$ (bottom) in Pb-Pb collisions at $\sqrt{s_{NN}}=2760$~GeV.}
\label{fig:directPhotonHadronicRates}        
\end{figure}

Calculations of direct photons in heavy ion collisions do not always use the same photon emission rates in the evaluation of thermal photons, which has a large impact on the level of agreement with data. The photon rates used in this work were summarized in Table~\ref{table:rates}. The contribution to photon emission of a $\pi$-$\rho$-$\omega$ system~\cite{Holt:2015cda} has just been published and was not included in previous calculations of direct photons. More importantly, parametrizations for the photon emission rate evaluated with the $\rho$ spectral function, along with additional emission from $\pi+\pi$ bremsstrahlung, were made available in Ref.~\cite{Heffernan:2014mla}. In consequence,  calculations of direct photons made before this point often included only photon emission from a meson gas. The importance of the different hadronic photon emission channels on photonic observables is shown in Fig.~\ref{fig:directPhotonHadronicRates}. Since the effects of shear and bulk viscosity have not yet been evaluated for all these photon emission rates, corrections to the photon rates due to viscosities are \emph{not} included for any emission rates in this comparison.

It is clear from Fig.~\ref{fig:directPhotonHadronicRates} that including only photon emission from a gas of mesons leads to a considerable underestimation of the direct photon $v_2$. The photon channels evaluated with the $\rho$ spectral function are especially important.

On a last note, it is relevant to highlight that QGP and hadronic photon emission rates that are altogether different from those used in the present work have been investigated over the past years. For completeness, the results of folding the hydrodynamical description of heavy ion collisions presented in this work with two of these rates are presented. The first rate is the ``semi-QGP'' photon emission rate~\cite{Gale:2014dfa}, which includes confinement effects on photon emission. The second rate is the hadronic rate from Zahed and Dusling~\cite{Dusling:2009ej}, which is evaluated using a different approach than the hadronic rates used in this work. Once again, viscous corrections to the photon rates are not included in this comparison.

\begin{figure}[tb]
        \includegraphics[width=0.35\textwidth]{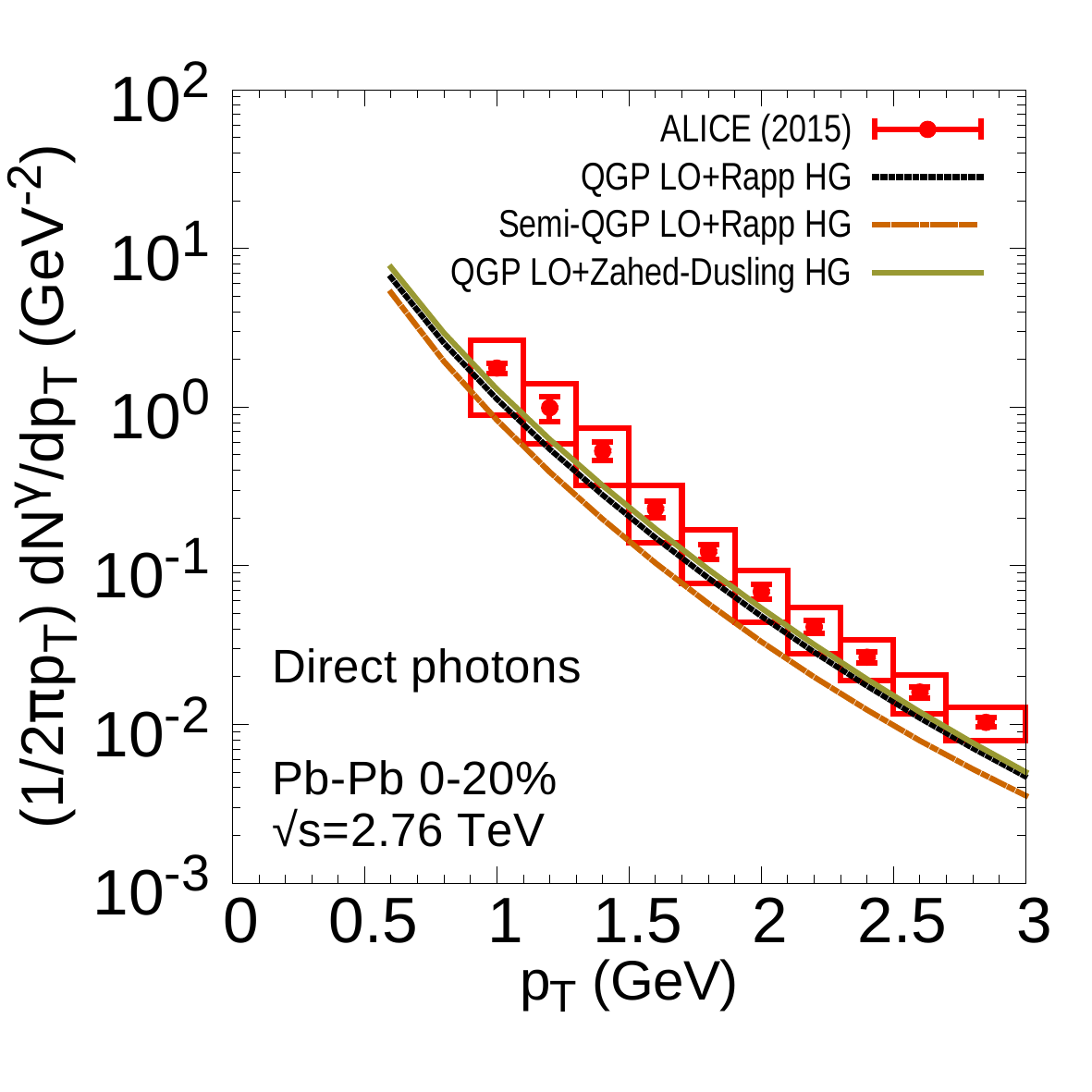}
        \includegraphics[width=0.35\textwidth]{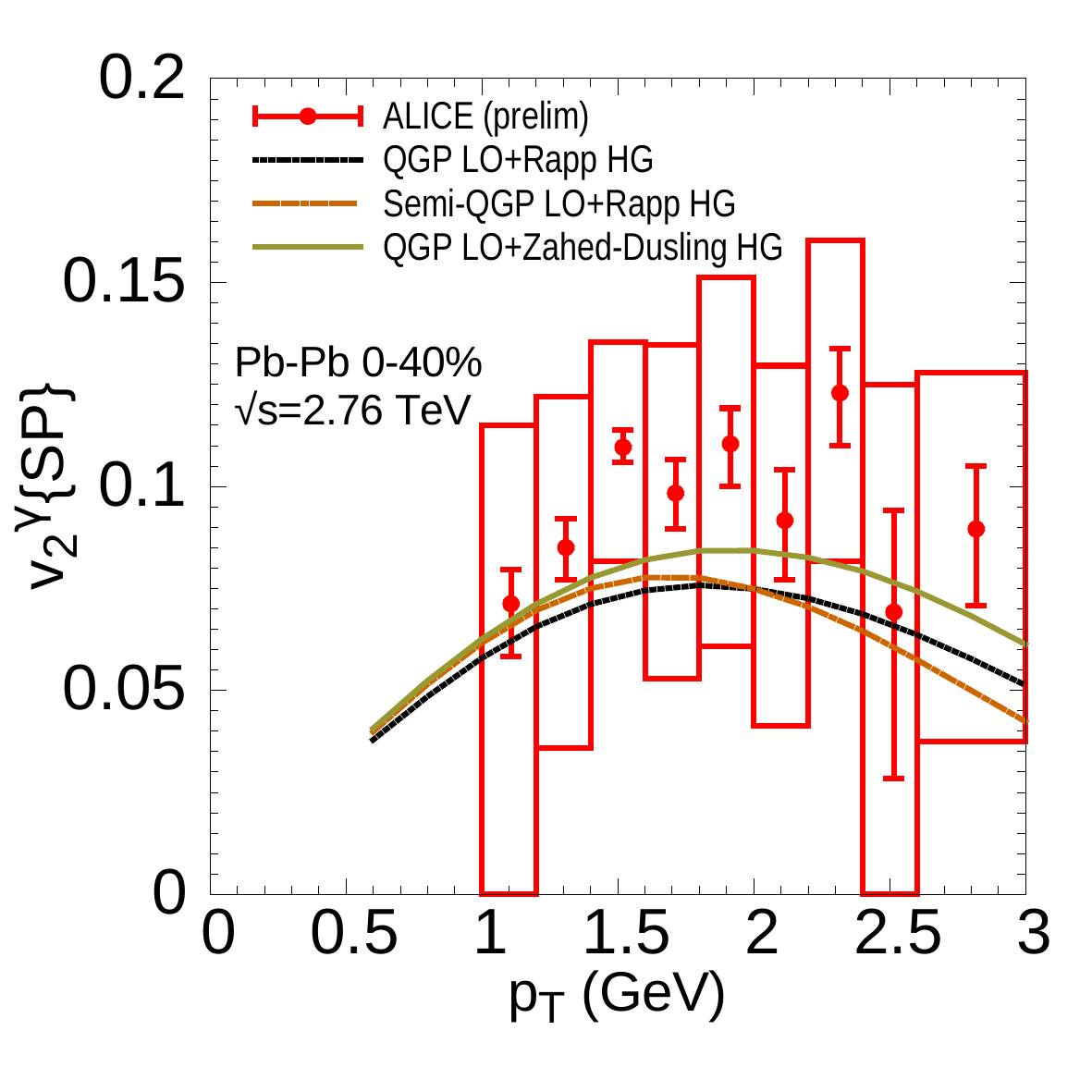}
\caption{Direct photon spectrum (top) and $v_2$ (bottom) evaluated with different QGP and hadronic photon emission rates, in Pb-Pb collisions at $\sqrt{s_{NN}}=2760$~GeV. See text for details.
}
\label{fig:directPhotonOtherRates}        
\end{figure}

The semi-QGP photon rate is considerably smaller than the QGP rate, which results in a 30\% suppression of the direct photon spectrum, shown on the top of Fig.~\ref{fig:directPhotonOtherRates}. The $v_2$, shown on the lower part of the figure, does not change significantly: an intuitive way of understanding this result is to note that while a suppression of the photon rate at high temperature will increase the \emph{thermal} photon $v_2$, it will reduce the contribution of thermal photons with respect to prompt photons. The two effects largely cancel out.

An important consequence of the suppression of the QGP rate studied in Ref.~\cite{Gale:2014dfa} is that it does not match well anymore the hadronic rate in the deconfinement region, as was the case with the QGP rate used previously in this work (Fig.~\ref{fig:rates}). This fact has yet to be addressed in a satisfactory fashion; it highlights the importance of understanding the photon rates in the transition region.

The hadronic rate from Ref.~\cite{Dusling:2009ej} is around 40\%-100\% larger than the one used in the present work. This results in a larger direct photon spectrum and a larger $v_2$, as illustrated in Fig.~\ref{fig:directPhotonOtherRates}. Since the hadronic rates being compared do not include the same photon production channels, this difference is not unexpected. Further studies of these hadronic emission rates themselves will be required to establish if the two approaches can be found to agree for comparable production channels.

\subsection{The importance of late stage photon emission}

\label{sec:latePhotons}
\begin{figure}[tb]
        \includegraphics[width=0.35\textwidth]{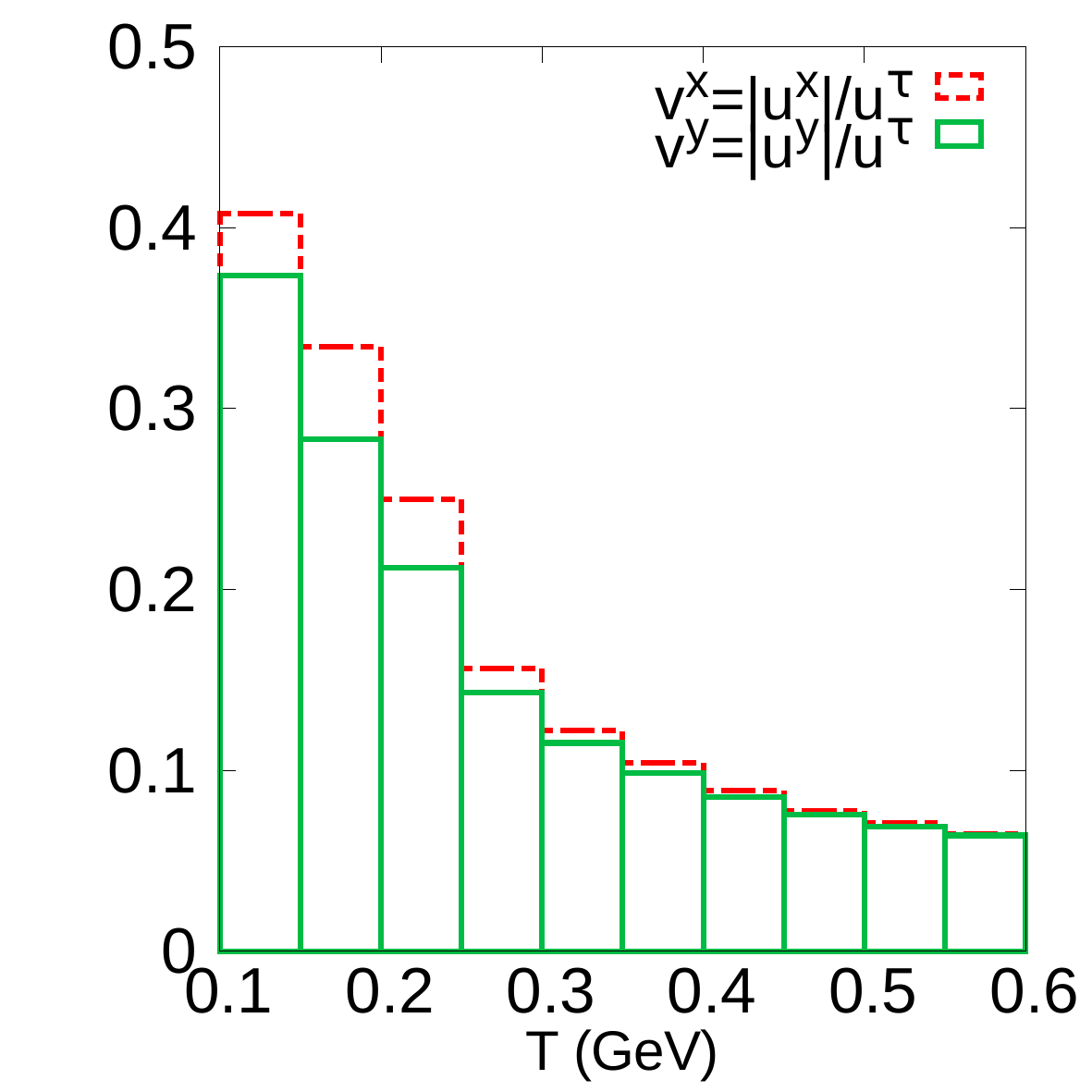}
        \includegraphics[width=0.35\textwidth]{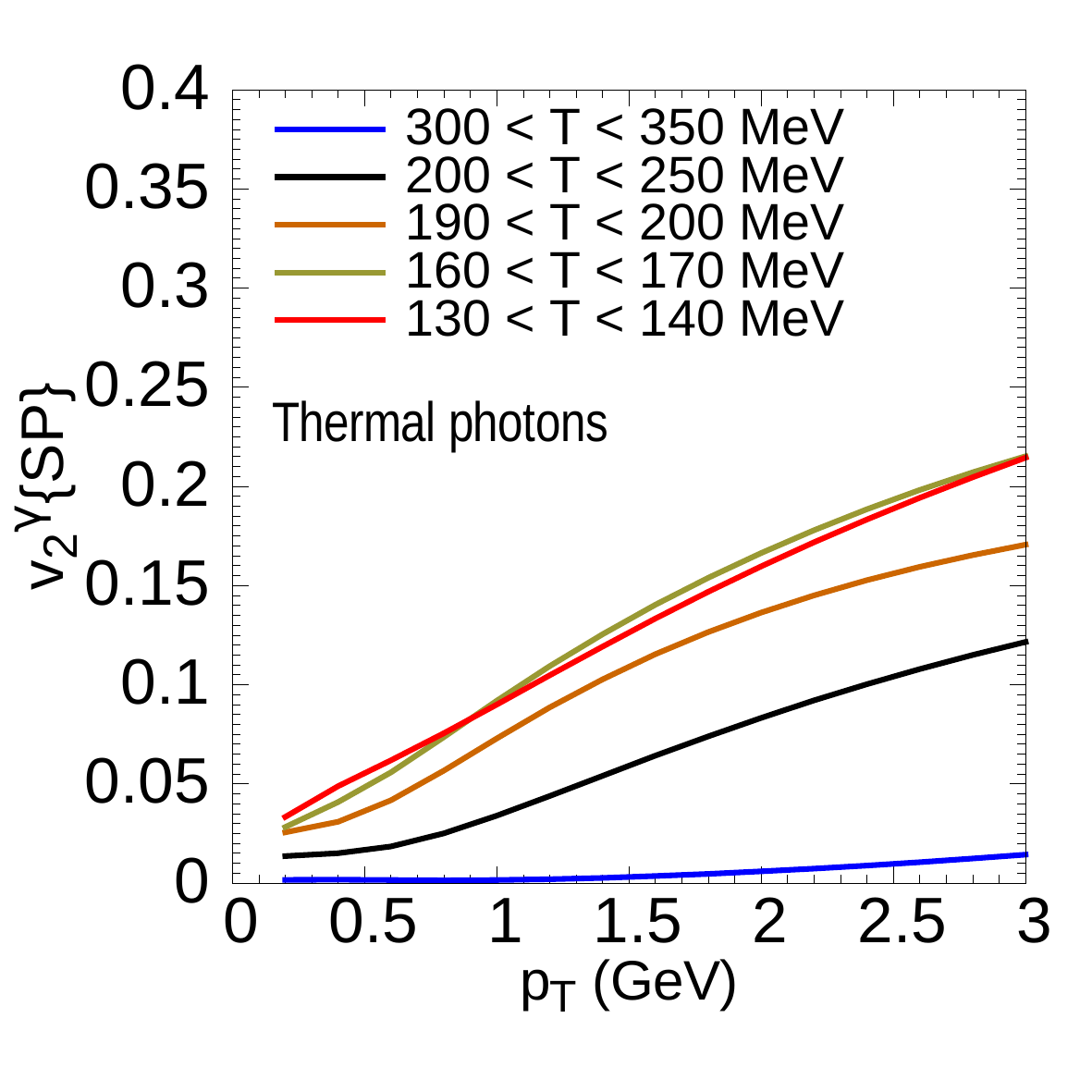}
\caption{Flow anisotropy of the medium with respect to the reaction plane (top) and momentum anisotropy of thermal photons (bottom), for different temperature ranges, in Pb-Pb collisions at $\sqrt{s_{NN}}=2760$~GeV.}
\label{fig:anisotropyVsT}        
\end{figure}

The current understanding of heavy ion collisions is that a flow velocity anisotropy is created during the medium's expansion as a result of the initial anisotropy in the energy deposition. This anisotropy increases with time, although it eventually plateaus and reverses under the effect of the viscosities and the decreasing pressure gradients. Using temperature as a proxy for time, this development of flow anisotropies is shown at the top of Fig.~\ref{fig:anisotropyVsT} as the $x$-$y$ flow asymmetry for different temperature ranges. The anisotropy is evaluated with respect to the reaction plane, with the $x$-axis aligned with the impact parameter of the colliding nuclei. Like hadrons, the momentum anisotropy of thermal photons is directly related to this flow velocity anisotropy. This is illustrated by plotting on the lower part of Fig.~\ref{fig:anisotropyVsT} the $v_2$ of thermal photons emitted in different regions of temperature. Lower temperatures are associated with larger time, which in turn is associated with larger thermal photon $v_2$.

\begin{figure}[tb]
        \includegraphics[width=0.35\textwidth]{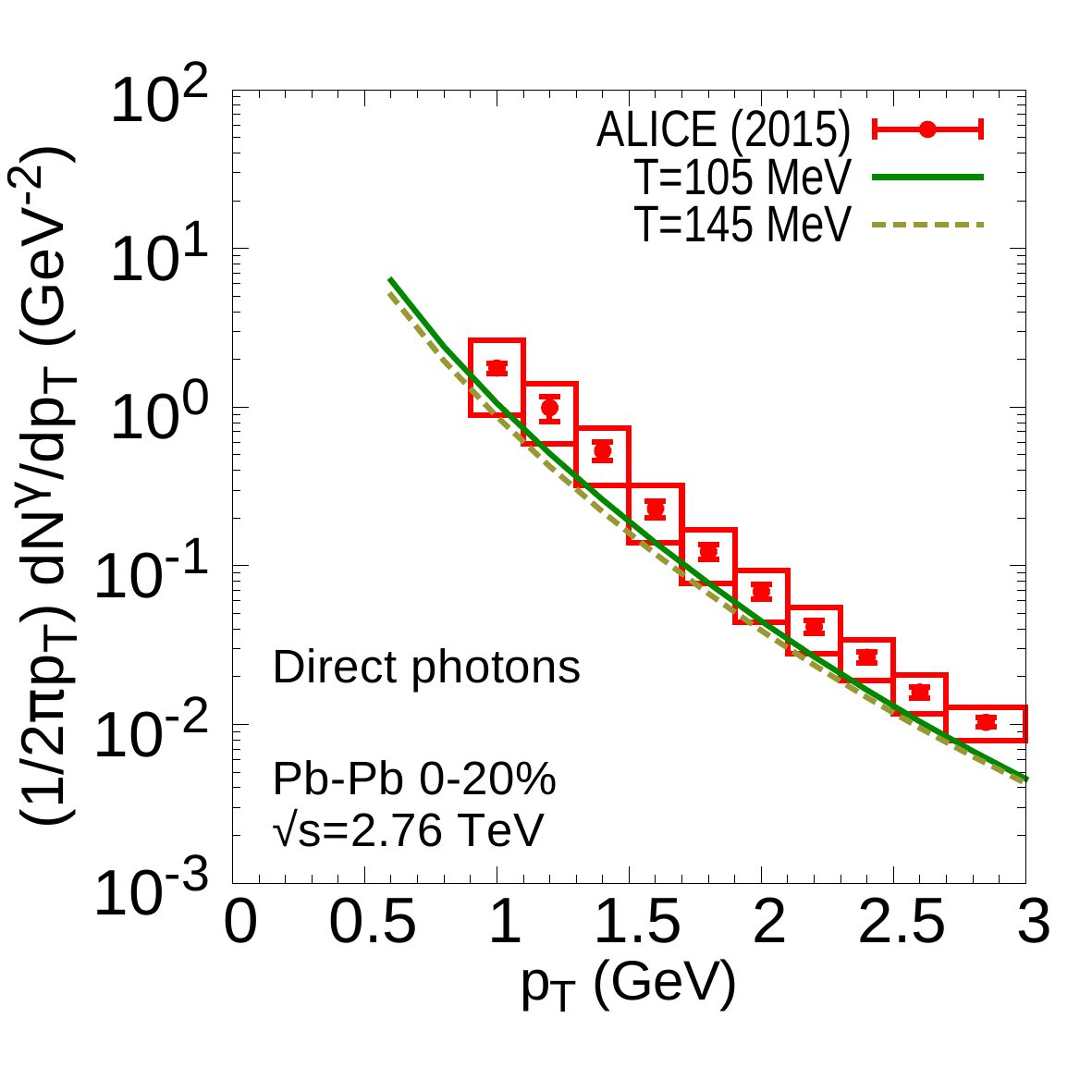}
        \includegraphics[width=0.35\textwidth]{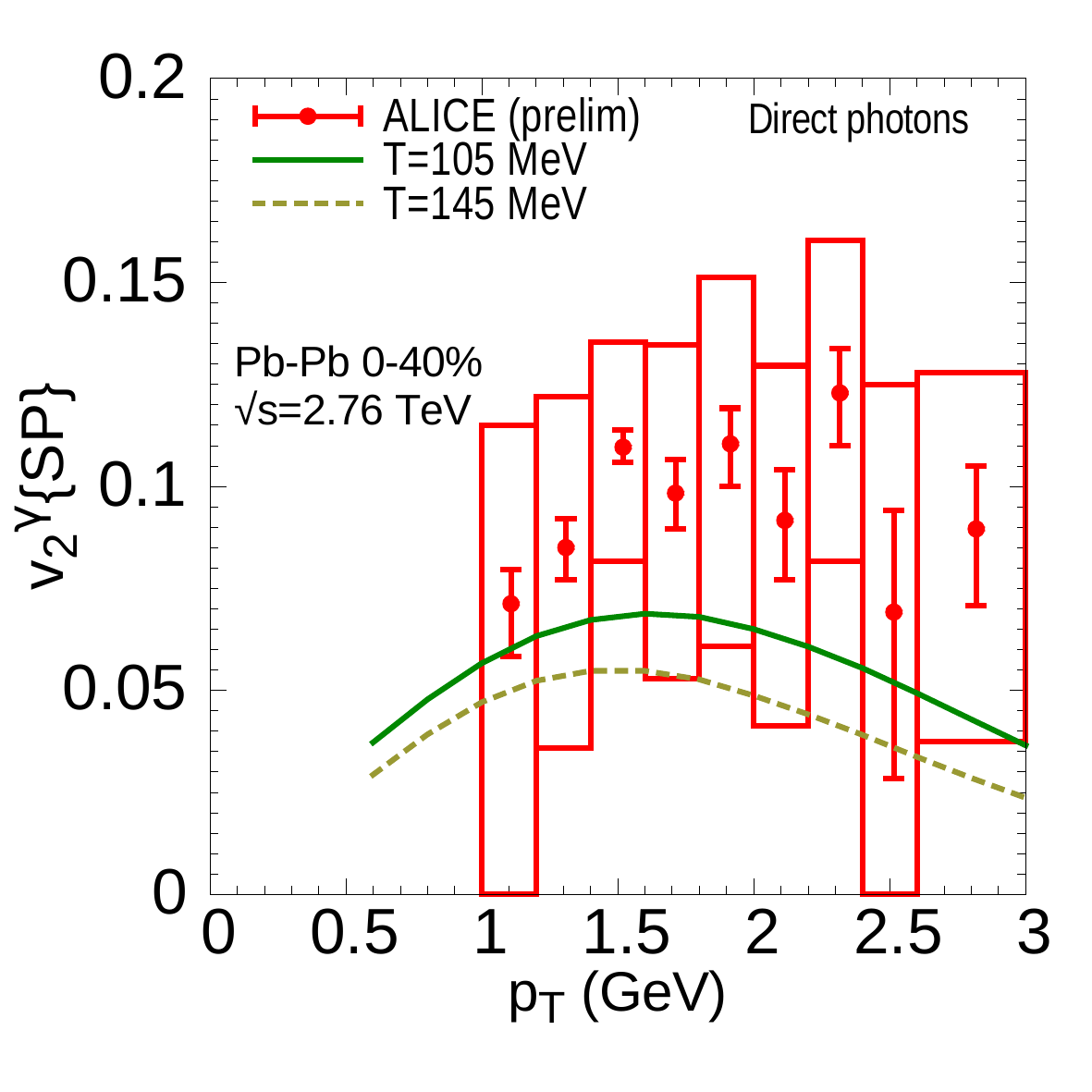}
\caption{Importance of post-particlization photon production on the photon spectrum (top) and $v_2$ (bottom) in Pb-Pb collisions at $\sqrt{s_{NN}}=2760$~GeV.}
\label{fig:directPhotonPostFO}        
\end{figure}

The magnitude of the $v_2$ of photons emitted at late times means that they can play a large role in increasing the direct photon $v_2$. Their importance is illustrated in Fig.~\ref{fig:directPhotonPostFO} by showing explicitly the contribution to the direct photon spectrum and $v_2$ of thermal photons emitted between $T=105$~MeV and $T=145$~MeV. This temperature range is chosen because $T=145$~MeV is the switching temperature between hydrodynamics and UrQMD that best describes the hadronic observables~\cite{Ryu:2015vwa} in Pb-Pb collisions at $\sqrt{s_{NN}}=2760$~GeV, as explained at the beginning of this section. Photons emitted below this switching temperature would thus be best evaluated using UrQMD~\cite{Bauchle:2010ym}, but this challenging task will be addressed in future work. The large contribution to the direct photon $v_2$ of these late stage photons highlights the importance of further studying photon emission during this phase of the evolution.

\section{Conclusions}

In this paper, the production of photons in heavy ion collisions was studied at RHIC and the LHC using a  hydrodynamical model of heavy ion collisions. This comprehensive model included realistic initial conditions (IP-Glasma), along with second-order hydrodynamics equations with both shear and bulk viscosities.  

With the inclusion of the elements comprising this paper, most direct photon theoretical results were found to lie either within the limits set by the systematic and statistical uncertainties of measurements at both colliders, or slightly below. A larger discrepancy  between theory and  PHENIX data remains, but in all cases the agreement between fluid dynamical calculations and experimental data were found to be improved compared to what they were in the past. This level of agreement with data provides strong support to the idea that thermal photons are the principal source of the low $p_T$ direct photon enhancement and of the large photon momentum anisotropy.

The presence of bulk viscosity in the hydrodynamical evolution produced a small effect on the photon spectrum at the LHC, and a modest change in the overall magnitude of the photon $v_2$. On the other hand, it induced a clear change in the shape of $v_2$, enhancing it at low $p_T^\gamma$ and reducing it at higher $p_T^\gamma$. While theoretical and experimental uncertainties do not currently permit to determine if this change in the shape of $v_2$ is favored by data, the reduction of both uncertainties in the future could allow direct photons to be used to constrain the bulk viscosity of QCD.

The significant contribution of late stage photon emission to the photon $v_2$, quantified in Section~\ref{sec:latePhotons}, highlights the need for a more sophisticated study of photon emission in this phase of the collisions. Work also remains to be done on constraining the photon emission rates, especially in the deconfinement region. These questions will be addressed in future work, and may help shed light on the different level of agreement observed with RHIC and LHC data.

Another topic necessitating further attention is the need for a more sophisticated treatment of prompt photon production in heavy ion collisions that includes both parton energy loss and jet-medium photon production \cite{Fries:2002kt,*Turbide:2005fk}. This could be especially important at RHIC, where thermal photons are not as dominant over prompt photons as at the LHC. The production of photons during the pre-thermalised phase of the collisions is also a part of the framework needing greater scrutiny. The encouraging results presented in this paper signal that the current understanding of thermal photons is mature enough for such investigations to be undertaken.

\begin{acknowledgements}
The authors would like to thank the organizers and participants of the ``EMMI Rapid Reaction Task Force on the direct-photon flow puzzle'', together with  Takao Sakaguchi, for fruitful discussions. The authors thank Kevin Dusling and Ismail Zahed for providing a tabulation of their photon emission rates. This work was supported in part by the Natural Sciences and Engineering Research Council of Canada. GSD and BPS are supported under DOE Contract No. DE-SC0012704. M.L.\ acknowledges support from the Marie Curie Intra-European Fellowship for Career Development grant FP7-PEOPLE-2013-IEF-626212.  Computations were made in part on the supercomputer Guillimin from McGill University, managed by Calcul Qu\'ebec and Compute Canada. The operation of this supercomputer is funded by the Canada Foundation for Innovation (CFI), NanoQu\'ebec, RMGA and the Fonds de recherche du Qu\'ebec - Nature et technologies (FRQ-NT). This research also used resources of the National Energy Research Scientific Computing Center, which is supported by the Office of Science of the U.S. Department of Energy under Contract No. DE- AC02-05CH11231. 

\end{acknowledgements}

\appendix

\section{Chapman-Enskog theory in the relaxation time approximation}

\label{appendixA}

The first-order Chapman Enskog approximation leads to the following formula
for the leading nonequilibrium correction to the distribution function, $%
\delta f_{i\mathbf{k}}$,%
\begin{equation}
Df_{i\mathbf{k}}^{\left( 0\right) }+\frac{1}{E_{\mathbf{k}}^{i}}k_{i}^{\mu
}\nabla _{\mu }f_{i\mathbf{k}}^{\left( 0\right) }=-\frac{\delta f_{i\mathbf{k%
}}}{\tau _{R}}\;,
\end{equation}%
where $D=u^{\mu} \partial_\mu$ and $\nabla^\mu = \Delta^{\mu\nu}
\partial_\mu $. The collision
term was already simplified using the relaxation time approximation $C\left[
\delta f_{i\mathbf{k}}\right] =-E_{\mathbf{k}}^{i}\delta f_{i\mathbf{k}%
}/\tau _{R}$, with $E_{\mathbf{k}}^{i}=u_{\mu }k_{i}^{\mu }$ and $\tau _{R}$
the relaxation time. We further defined the local equilibrium distribution
function $f_{i\mathbf{k}}^{\left( 0\right) }$, which corresponds to
Fermi-Dirac or Bose-Einstein distribution.

If the chemical potential and its derivatives are always set to zero, as
assumed in this paper, the derivatives of the local equilibrium distribution
function can be written as%
\begin{eqnarray}
\nabla _{\mu }f_{i\mathbf{k}}^{\left( 0\right) } &=&-\frac{1}{T}f_{0\mathbf{k%
}}\tilde{f}_{0\mathbf{k}}\left( -E_{\mathbf{k}}^{i}\nabla _{\mu }\ln
T+k_{i}^{\nu }\nabla _{\mu }u_{\nu }\,\right) , \\
Df_{i\mathbf{k}}^{\left( 0\right) } &=&-\frac{1}{T}f_{0\mathbf{k}}\tilde{f}%
_{0\mathbf{k}}\left( -E_{\mathbf{k}}^{i}D\ln T+k_{i}^{\nu }Du_{\nu }\right) ,
\end{eqnarray}%
and $\delta f_{i%
\mathbf{k}}$ can be expressed in the following way, 
\begin{eqnarray}
\frac{\delta f_{i\mathbf{k}}}{\tau _{R}} &=&\frac{1}{T}f_{0\mathbf{k}}\tilde{%
f}_{0\mathbf{k}}\left[ -E_{\mathbf{k}}^{i}D\ln T+k_{i}^{\nu }Du_{\nu
}\right.   \notag \\
&&\left. -k_{i}^{\mu }\nabla _{\mu }\ln T+\frac{1}{E_{\mathbf{k}}^{i}}%
k_{i}^{\mu }k_{i}^{\nu }\nabla _{\mu }u_{\nu }\,\right].
\end{eqnarray}

Then, using the well known thermodynamic relations (at vanishing chemical
potential)%
\begin{eqnarray*}
d\mathcal{P} &=&\frac{\epsilon +\mathcal{P}}{T}dT, \\
\frac{d\epsilon }{dT} &=&\frac{\epsilon +\mathcal{P}}{c_{s}^{2}T},
\end{eqnarray*}%
and the conservation laws, up to first order in Knudsen number,%
\begin{align}
D\epsilon & \approx -\left( \epsilon +\mathcal{P}\right) \theta , \\
\left( \epsilon +\mathcal{P}\right) Du^{\mu }& \approx \nabla ^{\mu }%
\mathcal{P}, \\
D\ln T& \approx -c_{s}^{2}\theta .
\end{align}%

We then obtain the following expression for $\delta f_{i\mathbf{k}}$ 
\begin{eqnarray*}
\frac{\delta f_{i\mathbf{k}}}{\tau _{R}} &=&\frac{1}{T}f_{0\mathbf{k}}\tilde{%
f}_{0\mathbf{k}}\left[ \left( c_{s}^{2}-\frac{1}{3}\right) E_{\mathbf{k}%
}^{i}\theta \right.  \\
&&\left. +\frac{1}{E_{\mathbf{k}}^{i}}\frac{m_{i}^{2}}{3}\theta +\frac{1}{E_{%
\mathbf{k}}^{i}}k_{i}^{\mu }k_{i}^{\nu }\sigma _{\mu \nu }\,\right].
\end{eqnarray*}%

The shear and bulk viscosity coefficients can be identified by replacing the
derived formula for $\delta f_{i\mathbf{k}}$ into the definitions for the
shear stress tensor, $\pi ^{\mu \nu }$, and the bulk viscous pressure, $\Pi $%
,%
\begin{eqnarray*}
\pi ^{\mu \nu } &=&\sum_{i=1}^{N}g_{i}\int dK_{i}\text{ }k_{i}^{\left\langle
\mu \right. }k_{i}^{\left. \nu \right\rangle }\delta f_{i\mathbf{k}}, \\
\Pi &=&-\frac{1}{3}\sum_{i=1}^{N}g_{i}\int dK_{i} \Delta_{\mu\nu} k_{i}^{\mu
}k_{i}^{\nu }\delta f_{i\mathbf{k}},
\end{eqnarray*}%
leading to the usual first order relations,%
\begin{eqnarray*}
\pi ^{\mu \nu } &=&2\hat{\eta}\tau _{R}\sigma ^{\mu \nu }, \\
\Pi &=&-\hat{\zeta}\tau _{R}\theta ,
\end{eqnarray*}%
where we introduced the transport coefficients $\hat{\eta}$ and $\hat{\zeta}$%
, 
\begin{eqnarray*}
\hat{\eta} &=&\frac{1}{5!! T}\sum_{i=1}^{N}g_{i}\int dK_{i}\text{ }%
f_{0\mathbf{k}}\tilde{f}_{0\mathbf{k}}\frac{1}{E_{\mathbf{k}}^{i}}\left(
\Delta _{\alpha \beta }k_{i}^{\alpha }k_{i}^{\beta }\right) ^{2}, \\
\hat{\zeta} &=&\frac{1}{3 T}\sum_{i=1}^{N}m_{i}^{2}g_{i}\int dK_{i}%
\text{ }f_{0\mathbf{k}}\tilde{f}_{0\mathbf{k}} \\
&&\qquad \qquad \times \left[ \left( c_{s}^{2}-\frac{1}{3}\right) E_{\mathbf{%
k}}^{i}+\frac{1}{E_{\mathbf{k}}^{i}}\frac{m_{i}^{2}}{3}\right] .
\end{eqnarray*}

Using the first order relations derived above, $\pi ^{\mu \nu }=2\hat{\eta}%
\tau _{R}\sigma ^{\mu \nu }$ and $\Pi =-\hat{\zeta}\tau _{R}\theta $, to
replace $\sigma ^{\mu \nu }$ and $\theta $ by $\pi ^{\mu \nu }$ and $\Pi $,
respectively, we obtain our final expression for $\delta f$, 
\begin{eqnarray}
\delta f_{i\mathbf{k}} &=&f_{0\mathbf{k}}\tilde{f}_{0\mathbf{k}}\frac{1}{2T%
\hat{\eta}E_{\mathbf{k}}^{i}}\pi _{\mu \nu }k_{i}^{\mu }k_{i}^{\nu }\, 
\notag \\
&&-\frac{1}{T\hat{\zeta}}f_{0\mathbf{k}}\tilde{f}_{0\mathbf{k}}\left[
-\left( \frac{1}{3}-c_{s}^{2}\right) E_{\mathbf{k}}^{i}+\frac{1}{E_{\mathbf{k%
}}^{i}}\frac{m_{i}^{2}}{3}\,\right] \Pi \;.  \notag \\
&&
\end{eqnarray}

Thus, the bulk correction to the single particle distribution function
becomes,

\begin{equation}
\delta f_{i\mathbf{k}}^{\Pi }\;=-\frac{1}{T\hat{\zeta}}\Pi f_{0\mathbf{k}}%
\tilde{f}_{0\mathbf{k}}\left[ \left( c_{s}^{2}-\frac{1}{3}\right) E_{\mathbf{%
k}}^{i}+\frac{m_{i}^{2}}{3E_{\mathbf{k}}^{i}}\,\right].
\end{equation}%
This is the expression used in the hadronic phase, both for photon and
hadron emission, with the index \textquotedblleft i\textquotedblright\
corresponding to a given hadron or resonance. When calculating $\hat{\zeta}$%
, all hadrons and resonances with masses up to 2.25 GeV were included.

\section{Chapman-Enskog theory in the relaxation time approximation with
thermal masses}

\label{appendixB}

Partons carry thermal masses. This changes some of the steps in the
derivation presented in the previous Appendix. In this case, we shall only keep the terms that contribute
to the bulk viscosity corrections to $\delta f_{i\mathbf{k}}$. We consider
the effective kinetic theory of quasiparticles derived in Ref.~\cite%
{Jeon:1995zm},%
\begin{equation*}
k^{\mu }_i\partial _{\mu }f_{i\mathbf{k}}-\frac{1}{2}\frac{\partial m_{i}^{2}}{%
\partial \mathbf{x}}\cdot \frac{\partial f_{i\mathbf{k}}}{\partial \mathbf{k}%
}=-E_{\mathbf{k}}^{i}\frac{\delta f_{i\mathbf{k}}}{\tau _{R}}.
\end{equation*}%
For our purposes, it is enough to assume that the mass goes with the
temperature as $m_{i}=g_{i}\left( T\right) T$. Then 
\begin{equation*}
k^{\mu }_i\partial _{\mu }f_{i\mathbf{k}}-m_{i}\frac{\partial m_{i}}{\partial 
\mathbf{x}}\cdot \frac{\partial f_{i\mathbf{k}}}{\partial \mathbf{k}}=-E_{%
\mathbf{k}}^{i}\frac{\delta f_{i\mathbf{k}}}{\tau _{R}}.
\end{equation*}

As before, the leading order solution originating from the Chapman-Enskog
approximation satisfies, 
\begin{equation*}
u_{\mu }k_{i}^{\mu }Df_{0\mathbf{k}}^{i}+k_{i}^{\mu }\nabla _{\mu }f_{0%
\mathbf{k}}^{i}-m_{i}\frac{\partial m_{i}}{\partial \mathbf{x}}\cdot \frac{%
\partial f_{0\mathbf{k}}^{i}}{\partial \mathbf{k}}=-E_{\mathbf{k}}^{i}\frac{%
\delta f_{i\mathbf{k}}}{\tau _{R}}.
\end{equation*}%
Since the mass depends on the temperature, the derivatives of the local
equilibrium distribution function have to be re-evaluated and one obtains
that (at vanishing chemical potential)%
\begin{eqnarray}
\nabla _{\mu }f_{i\mathbf{k}}^{\left( 0\right) } &=&-\frac{1}{T}f_{0\mathbf{k%
}}^{i}\tilde{f}_{0\mathbf{k}}^{i}\left( -E_{i\mathbf{k}}\nabla _{\mu }\ln
T+k_{i}^{\nu }\nabla _{\mu }u_{\nu }\right.   \notag \\
&&\left. \qquad \qquad \qquad \qquad +u_{\nu }\nabla _{\mu }k^{\nu
}\,\right) , \\
Df_{i\mathbf{k}}^{\left( 0\right) } &=&-\frac{1}{T}f_{0\mathbf{k}}^{i}\tilde{%
f}_{0\mathbf{k}}^{i}\left( -E_{i\mathbf{k}}D\ln T+k_{i}^{\nu }Du_{\nu
}\right.   \notag \\
&&\qquad \qquad \qquad \qquad \left. +u_{\nu }Dk^{\nu }\,\right) .
\end{eqnarray}%
In addition,%
\begin{equation*}
\frac{\partial f_{0\mathbf{k}}^{i}}{\partial \mathbf{k}}=-\frac{1}{T}f_{0%
\mathbf{k}}^{i}\tilde{f}_{0\mathbf{k}}^{i}u_{\mu }\frac{\partial k_{i}^{\mu }%
}{\partial \mathbf{k}} \;.
\end{equation*}

Using these results, we obtain the following expression for $\delta f_{i%
\mathbf{k}}$%
\begin{eqnarray}
&&\frac{1}{T}f_{0\mathbf{k}}^{i}\tilde{f}_{0\mathbf{k}}^{i}\left[ -E_{i%
\mathbf{k}}D\ln T+\frac{1}{3E_{i\mathbf{k}}}\Delta _{\alpha \beta }k^{\alpha
}_i k^{\beta }_i\theta +\frac{m_{i}}{E_{i\mathbf{k}}}\frac{\partial m_{i}}{%
\partial T}DT\right]   \notag \\
&=&\frac{\delta f_{i\mathbf{k}}^{\Pi }}{\tau _{R}} \;.
\end{eqnarray}%
Above, we only kept the terms which contribute to the bulk viscosity
correction, i.e., terms that are scalars in momentum space. Next, using the
conservation laws up to first order and thermodynamic relations, we can
further simplify this to%
\begin{eqnarray}
&&f_{0\mathbf{k}}^{i}\tilde{f}_{0\mathbf{k}}^{i}\left[ \left( E_{i\mathbf{k}%
}-\frac{m_{i}^{2}}{E_{i\mathbf{k}}}\frac{1}{g_{i}}\frac{\partial m_{i}}{%
\partial T}\right) c_{s}^{2}\qquad \qquad \right.   \notag \\
&&\qquad \qquad \qquad \qquad \left. +\frac{1}{3E_{i\mathbf{k}}}\left(
\Delta _{\alpha \beta }k^{\alpha }_i k^{\beta }_i\right) \right] \frac{\theta }{T}
\notag \\
&&\qquad \qquad =\frac{\delta f_{i\mathbf{k}}^{\Pi }}{\tau _{R}} \;.
\end{eqnarray}

Finally, using $\Pi =-\zeta \theta $, the bulk $\delta f$ can be written as%
\begin{eqnarray}
\delta f_{i\mathbf{p}} &=&-f_{0\mathbf{k}}^{i}\tilde{f}_{0\mathbf{k}}^{i}%
\left[ \left( \frac{m_{i}^{2}}{E_{i\mathbf{k}}}-E_{i\mathbf{k}}\right)
\left( \frac{1}{3}-c_{s}^{2}\right) \right.  \notag \\
&&\qquad \qquad \left. -\frac{\partial \ln g_{i}}{\partial \ln T}\frac{%
m_{i}^{2}}{E_{i\mathbf{k}}}c_{s}^{2}\right] \frac{1}{T} \frac{\Pi }{\zeta
/\tau _{\Pi }},
\end{eqnarray}%
where we used that $\tau _{R}=\tau _{\Pi }$, with $\tau _{\Pi }$ being the
bulk viscous pressure relaxation time.

For partonic degrees of freedom, the ratio $\frac{\zeta }{\tau _{\Pi }}$ is
approximated as 
\begin{equation*}
\frac{\zeta }{\tau _{\Pi }}=15\left( \epsilon +\mathcal{P}\right) \left( 
\frac{1}{3}-c_{s}^{2}\right) ^{2},
\end{equation*}%
which was derived in the quasi-conformal limit~\cite{Denicol:2014vaa}.

Using the above, we obtain%
\begin{eqnarray}
\delta  &&f_{i\mathbf{p}}=-f_{0\mathbf{k}}^{i}\tilde{f}_{0\mathbf{k}}^{i}%
\left[ \left( \frac{m_{i}^{2}}{E_{i\mathbf{k}}}-E_{i\mathbf{k}}\right)
\left( \frac{1}{3}-c_{s}^{2}\right) \right.   \notag \\
&&\,\,\left. -\frac{\partial \ln g_{i}}{\partial \ln T}\frac{m_{i}^{2}}{E_{i%
\mathbf{k}}}c_{s}^{2}\right] \frac{\Pi }{15T\left( \epsilon +\mathcal{P}%
\right) \left( \frac{1}{3}-c_{s}^{2}\right) ^{2}}\;.
\end{eqnarray}

If the running of $g_{s}$ is neglected, as is the case in the present work,
the expression reduces to 
\begin{eqnarray}
\delta f_{i\mathbf{p}} &=&-f_{0\mathbf{k}}^{i}\tilde{f}_{0\mathbf{k}}^{i}%
\left[ \left( \frac{m_{i}^{2}}{E_{i\mathbf{k}}}-E_{i\mathbf{k}}\right)
\left( \frac{1}{3}-c_{s}^{2}\right) \right]   \notag \\
&&\times \frac{\Pi }{15T\left( \epsilon +\mathcal{P}\right) \left( \frac{1}{3%
}-c_{s}^{2}\right) ^{2}}\;.
\end{eqnarray}

We note that for a quark, anti-quarks, and gluons, the values of the
asymptotic masses are%
\begin{eqnarray*}
m_{q,\bar{q}}^{2} &=&\frac{g^{2}T^{2}}{3} \\
m_{g}^{2} &=&\left( 3+\frac{N_{F}}{2}\right) \frac{g^{2}T^{2}}{6} \;.
\end{eqnarray*}

\bibliography{biblio}

\end{document}